\documentclass[onecolumn]{autart}    

\usepackage{graphicx}          

\usepackage{cite}
\usepackage{bm}
\usepackage{amsmath}
\usepackage{graphicx}
\usepackage{amssymb}
\usepackage{multirow}
\usepackage{array}
\usepackage{color}
\usepackage{epsfig}
\usepackage{algorithmic}
\usepackage{graphics} 
\usepackage{epsfig} 
\usepackage{subcaption}

\newcommand{\black}{\color{black}}

\newcommand{\tcS}{\tilde{\mathcal{S}}}

\allowdisplaybreaks

\newtheorem{definition}{Definition}
\newtheorem{theorem}{Theorem}
\newtheorem{proposition}{Proposition}

\newtheorem{assumption}{Assumption}

\newtheorem{example}{Example}

\newcommand{\cS}{\mathcal{S}}

\def\bull{\vrule height 1.1ex width 1.1ex depth -.0ex }

\begin{document}

\begin{frontmatter}

\title{Coordination Control of Discrete Event Systems under Cyber Attacks\thanksref{footnoteinfo}} 

\thanks[footnoteinfo]{This work is supported in part by the National Natural Science Foundation of China under Grant 62073242, and by RVO 67985840. Corresponding author Feng Lin. Tel. +313-5773428.}

\author[Paestum]{Fei Wang}\ead{feiwang@sit.edu.cn},    
\author[Rome]{Jan Komenda}\ead{komenda{@}ipm.cz},               
\author[Baiae]{Feng Lin}\ead{flin@wayne.edu}  

\address[Paestum]{School of Electrical and Electronic Engineering, Shanghai Institute of Technology, Shanghai 201416, China.}  
\address[Rome]{Institute of Mathematics, Academy of Sciences of the Czech Republic, {\v Z}itn\'a 25, 115 67 Prague, Czech Republic.}             
\address[Baiae]{Department of Electrical and Computer Engineering, Wayne State University, Detroit, MI 48202, USA.}        


\begin{abstract}                          
{\black In this paper, coordination control of discrete event systems under joint sensor and actuator attacks is investigated. Sensor attacks are described by a set of attack languages using a proposed ALTER model.
	Several local supervisors are used to control the system. The goal is to design local supervisors to ensure safety of the system even under cyber attacks (CA).
	The necessary and sufficient conditions for the existence of such supervisors are derived in terms of conditional decomposability, CA-controllability and CA-observability. A method is developed to calculate local state estimates under sensor attacks.
	Two methods are also developed to design local supervisors, one for discrete event systems satisfying conditional decomposability, CA-controllability and CA-observability, and one for discrete event systems satisfying conditional decomposability only. The approach works for both stealthy and non-stealthy attacks.}
	{\black Two examples are given to illustrate the results.}
\end{abstract}

\end{frontmatter}

\section{Introduction}

Supervisory control of discrete event systems (DES) initially investigates how to control a system (plant) $G$ using one (global) supervisor $\cS$. The necessary and sufficient conditions for the existence of a supervisor $\cS$ which generates a given specification language $K$ are characterized by controllability \cite{ramadge1987supervisory} and observability \cite{lin1988observability} of $K$.
This (centralized) supervisory control has been extended to decentralized control, modular control, and  coordination control \cite{rudie1992think, wong1998modular, komenda2008coordination, queiroz2000modular, komenda2012supervisory}.

In coordination control, we start with $G$ and $K$. We design two or more local supervisors based on local versions of $G$ and $K$. To do so, we properly select two or more sets of local events to ensure that $K$ satisfies two conditions.
The first condition is conditional decomposability \cite{komenda2008coordination}, which says that the synchronous product of the projections of $K$ {\black to some sets of local events} is equal to $K$. The second condition is the observer property \cite{pena2008polynomial}, which says that the numbers of states in {\black the projected automata} is smaller than the number of states in $G$.
The two conditions can always be satisfied by properly augmenting the local alphabets (event sets). When they are satisfied, coordination control is highly effective for supervisory control of DES.
It simplifies the design, improves implementation, and ensures scalability while reducing the computational complexity \cite{komenda2008coordination, komenda2012supervisory, komenda2015distributed}. In this paper, we use coordination control for discrete event system under cyber attacks (CA).

As networks become more and more widely used, supervisory control theory encounters new challenges.
One challenge is how to handle cyber attacks. It is well-known that cyber attacks are very serious threats to engineering systems.
{\black Researchers have investigated supervisory control of DES under cyber attacks, for example, in \cite{fritz2018modeling, wakaiki2019supervisory, zhang2018stealthy, wang2019supervisory,  carvalho2016detection, wang2020mitigation, alves2022discrete, zhang2021joint, ma2022resilient, tai2023synthesis, meira2020moving, lima2021security}.
A good overview of the works on cyber attacks can be found in \cite{rashidinejad2019supervisory}. However, the problem of coordination control for discrete event system under cyber attacks remains largely unsolved.
}

{\black We will investigate coordination control for discrete event system under sensor attacks using the ALTER model (for Attack Language, Transition-basEd, Replacement) proposed \cite{zheng2021modeling, zheng2023modeling}.
The ALTER model is general and can be used to model a large class of cyber attacks, including deletions, insertions, replacements, and all-out attacks.
Using the ALTER model, an attacker can replace the event in an attackable transition by any one of the strings in the corresponding attack language.
Attack languages are application specific and can be obtained based on the designer's knowledge of the attacks. For example, if the designer has no knowledge of the attacker (as in all-out attacks), then the attack languages will be large and include all possible attacks.
One consequence of large attack languages is that existence condition of (local) supervisors obtained in this paper is less likely to be satisfied.
The designer can then either try to learn more about the attackers (and hence obtain smaller attack languages), or the designer can design more restrictive supervisors that achieve a sublanguage of $K$. The choice is left for the designer to make.
Therefore, the ALTER model can handle a range of situations, where the knowledge of the attacker varies. As we will shown in this paper, knowing more about the attacker can improve the control.

For cyber attacks on actuators, we assume that an attacker can alter control commands issued by a supervisor by changing the disablements and enablements of attackable events.	

Due to nondeterminism in sensor attacks and actuator attacks, the language generated by the supervised system is not unique.
An upper bound on the set of languages generated by the supervised system, called large language, will be used to ensure safety, that is, the large language is equal to or contained in $K$.

For coordination control, we define the large language of the overall system and the large languages of the local subsystems and show the relation between them. We then derive the necessary and sufficient conditions for the existence of local supervisors capable of enforcing the global system specification under cyber attacks.
When the conditions are met, we propose systematic design procedures for constructing local supervisors that mitigate the effects of cyber attacks and achieve $K$.
The supervisors are based on state estimates. Therefore, a method is developed to calculate local state estimates under sensor attacks.

For cases where the conditions are not satisfied, we propose systematic design procedures for constructing local supervisors that mitigate the effects of cyber attacks and achieve a sublanguage of $K$.
Our work provides a comprehensive approach for designing resilient and robust supervisory systems, ensuring reliable operation in adversarial environments. Furthermore, our approach works for both stealthy and non-stealthy attacks, which makes the approach more useful because both stealthy and non-stealthy attacks occur in reality. In the literature, an attack is stealthy if an observer or supervisor can not detect it, that is, the attacked system produces only observations that belong to the observed language as the supervisor sees it. If the supervisor knows that the attacker is stealthy, then it gives the supervisor more information about the attacker. Hence, the attack languages of the ALTER model can be reduced, as some strings in the attack languages are not stealthy and should be removed. Smaller attack languages increase the likelihood that the existence condition for supervisor is satisfied, which then leads to a more permissive closed-loop.
}

The paper is organized as follows. In Section 2, we introduce the necessary notations and review supervisory control of DES under cyber attacks proposed in \cite{zheng2023modeling}.
In Section 3, we investigate coordination control of DES under cyber attacks. We derive the necessary and sufficient conditions for the existence of local supervisors resilient to cyber attacks.
In Section 4, we propose a design procedure for local resilient  supervisors when the necessary and sufficient conditions are satisfied.
In Section 5, we propose a design procedure for local resilient supervisors when conditionally decomposability is satisfied only.
In Section 6, an illustrative example is given to show how to handle two different attacks; a second example is provided to show how to apply the theoretical results to practical systems.

\section{Supervisory Control of Discrete Event Systems under Cyber Attacks}

Let us review discrete event systems, supervisory control, and cyber attacks in this section.

\subsection{Discrete Event Systems and Supervisory Control}

As usual, we model a discrete event system by a finite deterministic automaton \cite{cassandras2021introduction, wonham2019supervisory}:
$$
G=( Q, \Sigma , \delta , q_o, Q_m ),
$$
where $Q$ is a finite set of states; $\Sigma$ is a finite set of events; $\delta : Q \times \Sigma \rightarrow Q$ is the (partial) transition function;
$q_o$ is the initial state; and $Q_m\subseteq Q$ is the set of marked states. With a slight abuse of notation, the set of all possible transitions is also denoted by $\delta$ as $\delta = \{ (q, \sigma, q'): \delta (q, \sigma) =q' \}$.

We use $\Sigma ^*$ to denote the set of all (finite) strings over $\Sigma$, including the empty string $\varepsilon$. The transition function $\delta$ on single event $\sigma$ is extended to strings $s$, that is, $\delta : Q \times \Sigma^* \rightarrow Q$ in the usual way \cite{cassandras2021introduction, wonham2019supervisory}.
If $\delta (q,s)$ is defined, we denote it by $\delta (q,s)!$. The language {\em generated} by $G$ is the set of all strings defined in $G$ from the initial state:
$$
L(G) = \{ s \in \Sigma ^*: \delta (q_o,s)! \}.
$$

Similarly, the language {\em marked} by $G$ is the set of all strings  from the initial state that end up in a marked state of $G$:
$$
L_m(G) = \{ s \in \Sigma ^*: \delta (q_o,s)\in Q_m \}.
$$

In general, a language $L \subseteq \Sigma ^*$ is a set of strings. For $s\in \Sigma^*$, $s'\in \Sigma^*$ is its prefix if there exists $s''\in \Sigma^*$  such that $s=s's''$. The (prefix) closure of $L$ is the set of prefixes of strings in $L$.
It is denoted by $\overline{L}$. A language is (prefix) closed if it equals its prefix closure. By definition, $L(G)$ is closed.
The length of a string $s \in \Sigma ^*$ is denoted by $|s|$. The cardinality (the number of its elements) of a set $x \subseteq Q$ is denoted by $|x|$.
For a language $L$, the number of states of the minimal generator of $L$ (automaton $G$ generating $L=L(G)$) is denoted by $\| L \| $.

The system $G$ is controlled by a supervisor $\cS$, which can observe some observable events and control (disable) some controllable events.
The set of observable events is denoted by $\Sigma _o$ and the set of unobservable events is denoted by $\Sigma_{uo} = \Sigma\setminus\Sigma_o$.
The set of controllable events is denoted by $\Sigma _c$ and the set of uncontrollable events is denoted by $\Sigma_{uc} = \Sigma\setminus\Sigma_c$.

The observation of a supervisor is described by the natural projection $P: \Sigma ^* \rightarrow \Sigma _o^*$ defined as
\begin{align*}
	P(\varepsilon)= & \varepsilon, &
	P(s \sigma)= & \begin{cases}
		P(s) \sigma &\text{if $\sigma \in \Sigma_o$}\\
		P(s) &\text{if $\sigma\in \Sigma_{uo}$}\\
	\end{cases}
\end{align*}
where $\varepsilon$ is the empty string.
The inverse mapping of $P$ is defined as $P^{-1} (w) = \{ s \in \Sigma ^* : P(s)=w \}$. $P$ and $P^{-1}$ are extended to a language in the usual way.

For modular control with $n$ local sites, we consider local alphabets (event sets) $\Sigma _i, i=1,\ldots,n,$ with $\Sigma= \cup_{i=1}^{n} \Sigma_i$ and local automata $G_i  = (Q_i, \Sigma_i , \delta _i , q_{i,o})$  over alphabets $\Sigma_i$.
Recall that {\em synchronous product\/} of languages $L_i\subseteq \Sigma_i^*$ is the language
$$
\|_{i=1}^{n} L_i = \cap_{i=1}^{n} P_i^{-1}(L_i)\subseteq \Sigma^*,
$$
where $P_i \colon \Sigma^* \to \Sigma_i^*$  are projections to local alphabets.
A corresponding definition of synchronous product $\|_{i=1}^{n} G_i$ can also be found in \cite{cassandras2021introduction}. We have
\begin{align*}
	L(\|_{i=1}^{n} G_i) = \|_{i=1}^{n} L(G_i),
	L_m(\|_{i=1}^{n} G_i) = \|_{i=1}^{n} L_m(G_i) .
\end{align*}

\subsection{Supervisory Control under Cyber Attacks}

We consider both sensor attacks (attacks in an observation channel from a plant to a supervisor) and actuator attacks (attacks in a control channel from a supervisor to a plant).
{For sensor attacks, we use the ALTER model proposed in \cite{zheng2023modeling}, which is reviewed as follows.} Denote the set of (observable) events and transitions that may be attacked by $\Sigma _o ^a \subseteq \Sigma _o$ and $\delta ^a = \{ (q, \sigma, q') \in \delta : \sigma \in \Sigma _o ^a \}$, respectively. We call events $\sigma \in \Sigma_o ^a$ attackable events and transitions ${tr}=(q, \sigma, q')\in \delta ^a$ attackable transitions.

An attacker can observe observable events. When $\sigma \in \Sigma_o^a$ in ${tr}=(q, \sigma, q')\in \delta ^a$ occurred in $G$ and is intercepted by the attacker, the attacker can change the event $\sigma$ to any string in an attack language $A_{\sigma} \subseteq \Sigma ^*$.
Note that, since unobservable events neither help nor confuse the supervisor, whether we require $A_{\sigma} \subseteq \Sigma ^*$ or $A_{\sigma} \subseteq \Sigma _o ^*$ will make no difference.

Hence, if a string $s = \sigma _1 \sigma _2 ... \sigma _{|s|} \in L(G)$ occurs in $G$, the set of possible strings after sensor attacks, denoted by $\Theta^a (s)$.
$
\Theta^a (s) = A_{\sigma_1} A_{\sigma_2} ... A_{\sigma_{|s|}},
$

This sensor attack model is general and includes the following special cases.
(1) No attack: if $\sigma \in A_{\sigma}$ and $\sigma$ is altered to $\sigma$ (no change), then there is no attack. That is $A_{\sigma}=\{ \sigma,\cdots\}$.
(2) Deletion: if the empty string $\varepsilon \in A_{\sigma}$ and $\sigma$ is altered to $\varepsilon$, then $\sigma$ is deleted. That is $A_{\sigma}=\{ \varepsilon,\cdots\}$.
(3) Replacement: if $\alpha \in A_{\sigma}$ and $\sigma$ is altered to $\alpha$, then $\sigma$ is replaced by $\alpha$. That is $A_{\sigma}=\{ \alpha,\cdots\}$.
(4) Insertion: if $\sigma \alpha \in A_{\sigma}$ (or $\alpha \sigma \in A_{\sigma}$) and $\sigma$ is altered to $\sigma \alpha$ (or $\alpha \sigma$), then $\alpha$ is inserted. That is $A_{\sigma}=\{ \sigma \alpha,\cdots\}$.
(5) All-out attack: if $A_{\sigma}=\Sigma^*$, then the attacker can change $\sigma$ to any strings.Note that, from practical viewpoint, we may want to assume that $A_{\sigma}$ is finite, although such an assumption is not needed theoretically.

The resulting observation under both partial observation and sensor
attacks is then given by, for $s \in L(G)$,
\begin{equation} \label{Equation3}
	\begin{split}
		\Phi^a (s) = P \circ \Theta^a  (s) = P(\Theta^a  (s)) .
	\end{split}
\end{equation}
$\Theta^a $ and $\Phi ^a$ are extended from strings $s$ to languages $K \subseteq L(G)$ in the usual way as
\begin{equation} \label{Equation4}
	\begin{split}
		& \Theta^a (K) = \{ w \in \Sigma^*: (\exists s \in K) w  \in
		\Theta^a (s) \}, \\
		& \Phi ^a(K) = \{ w \in \Sigma^*_o: (\exists s \in K) w  \in \Phi
		^a(s) \}.
	\end{split}
\end{equation}

To model sensor  attacks for automaton $G$, we do the following. For each attackable event $\sigma \in \Sigma_o^a$, let $A_{\sigma}$ be marked by an automaton $F_{\sigma}$, that is, $A_{\sigma} = L_m (F_{\sigma})$ for some
$$
F_{\sigma} = (Q_{\sigma} , \Sigma, \delta_{\sigma}, q_{0,{\sigma}}, Q_{m,{\sigma}}).
$$
{Note that not all states in $F_{\sigma}$ are marked. We assume that $F_{\sigma}$ is trim, that is, all states are accessible from $q_{o,{\sigma}}$ and co-accessible to $Q_{m,{\sigma}}$.}

Let us first show how to substitute an attackable transition ${tr}=(q, \sigma, q') \in \delta ^a$ by a set of sequences of transitions $((q,\varepsilon, q_{0,\sigma}), \delta_{\sigma}, (q_{m,\sigma},\varepsilon,q'))$ (with $\delta_{\sigma}$ being the set of transition sequences defined in $F_{\sigma}$). We use the intuitive notation $G_{{ (q, \sigma, q')} \rightarrow (q, F_{\sigma}, q')}$ to denote this substitution procedure, which is formalized as follows.
\begin{align*}
	G_{{ (q, \sigma, q')} \rightarrow (q, F_{\sigma}, q')} = ( Q \cup Q_{\sigma}, \Sigma ,
	\delta_{{ (q, \sigma, q')} \rightarrow (q, F_{\sigma}, q')}, q_o ,{Q_m}),
\end{align*}
where $\delta_{ (q, \sigma, q') \rightarrow (q, F_{\sigma}, q')} = (\delta\setminus\{ (q, \sigma, q') \} ) \cup \delta_{\sigma} \cup \{ (q, \varepsilon, q_{0,\sigma}) \}$  $\cup \{ (q_{m,\sigma}, \varepsilon, q'): q_{m,\sigma} \in Q_{m,\sigma} \}$.
In other words, $G_{(q, \sigma, q') \rightarrow (q, F_{\sigma}, q')}$ contains all transitions in $\delta$ and $\delta_{\sigma}$, except $(q, \sigma, q')$, plus the $\varepsilon$-transitions from $q$ to $q_{0,{\sigma}}$ and from $Q_{m,{\sigma}}$ to $q'$.
{If $F_{\sigma}$ has only one marked state, that is, $Q_{m,\sigma}=\{ q_{m,\sigma} \}$, then we can use a shortcut by merging $q$ and $q'$ with $q_{0,\sigma}$ and $q_{m,\sigma}$, respectively, 	without the $\varepsilon$-transitions $(q, \varepsilon, q_{0,\sigma})$ and $(q_{m,\sigma}, \varepsilon, q')$.
}

The extended automaton obtained after replacing all
transitions subject to attacks is denoted by
\begin{equation} \label{eq:Ga}
	\begin{split}
		G^a =( Q^a, \Sigma , \delta ^a , q_o, Q^a_m)
		=( Q \cup \tilde{Q}, \Sigma , \delta ^a , q_o, Q),
	\end{split}
\end{equation}
where $\tilde{Q}= \cup_{\sigma \in \delta ^a} Q_{\sigma}$ is the set of states added during the substitution and $Q^a_m=Q$ is the set of marked states.
$G^a$ models the attacked system by combining $G$ and the ALTER model.
Note that $G^a$ is a nondeterministic automaton and its transition is a mapping $\delta ^a: Q^a \times \Sigma \rightarrow 2^{Q^a}$. From the construction of $G^a$, it can be shown \cite{zheng2023modeling} that
\begin{equation} \label{LmGe}
	\begin{split}
		 L_m(G^a) = \Theta^a (L(G)),
		 L(G^a) = \overline{\Theta^a (L(G))} .
	\end{split}
\end{equation}

To describe the partial observation, we replace unobservable
transitions in $G^a$ by $\varepsilon$-transitions and denote the
resulting automaton as
\begin{equation} \label{eq:Gaeps}
	\begin{split}
		G_{\varepsilon}^a &= ( Q^a, \Sigma _{o,\varepsilon} , \delta _{\varepsilon}^a ,
		q_o, Q^a_m)
		= ( Q \cup \tilde{Q}, \Sigma _{o,\varepsilon} , \delta
		_{\varepsilon}^a , q_o, Q),
	\end{split}
\end{equation}
where $\Sigma _{o,\varepsilon}=\Sigma _o \cup \{\varepsilon\}$ and $\delta _{\varepsilon}^a = \{ (q, \sigma, q'): (q, \sigma,
q') \in \delta ^a \wedge \sigma \in \Sigma _o \} \cup \{ (q,
\varepsilon, q'): (q, \sigma, q') \in \delta ^a \wedge \sigma
\not\in \Sigma _o \} $. Clearly, $G_{\varepsilon}^a$ is a
nondeterministic automaton.

$G_{\varepsilon}^a$ marks the language $\Phi^a (L(G))$ because
$$
L_m(G_{\varepsilon}^a) = P(L_m(G^a)) = P(\Theta^a  (L(G))) = \Phi^a (L(G)).
$$

Sensor attacks can be either ``static'' or ``dynamic''. A sensor attack is static if, for any two transitions ${tr_1}=(q_1, \sigma, q_1') \in \delta ^a$ and ${tr_2}=(q_2, \sigma, q_2') \in \delta ^a$ with the same event $\sigma$, we have $A_{tr_1} = A_{tr_2}$.
Hence, $A_{\sigma}$ can also be written as $A_\sigma$.
A sensor attack is dynamic if the above is not true, that is, $A_{tr_1} \not= A_{tr_2}$ for some ${tr_1}=(q_1, \sigma, q_1') \in \delta ^a$ and ${tr_2}=(q_2, \sigma, q_2') \in \delta ^a$ with the same event $\sigma$.

For static attacks, the implementation is  simple:  whenever an attacker sees an event $\sigma \in \Sigma _o ^a$, it will replace every $\sigma$ with some strings in the same $A_\sigma$. In this case, the plant model $G$ can be used as is.

For dynamic attacks, the implementation is more complex: when an attacker sees an event $\sigma \in \Sigma _o ^a$, it has the option of choosing a different attack language to use, since this language may not be the same for each occurrence of $\sigma$. We assume that the attacker's decision depends on the string of events it has observed so far.

Formally, an implementation of dynamic attacks is based on a finite automaton:
$$
M=( Y, \Sigma _{o} , \zeta , y_o , Y_m).
$$
We assume that all states in $M$ are marked, that is, $Y_m=Y$. To ensure that all strings observed by the attacker are defined in $M$, it is required that $P(L(G)) \subseteq L(M)$.
For example, we can use the observer of $G$, $P(G)$, as $M$. In that case, $P(L(G)) = L(M)$.
Note that $M$ is just an auxiliary automaton that aims to distinguish(using synchronous composition $\hat G$ below) two or more transitions sharing the same event so that the attacker  applies different (dynamic) attack languages to these transitions.

The set of all possible transitions of $M$ is denoted by $\zeta = \{ (y, \sigma, y'): \zeta (y, \sigma) =y' \}$. The set of attackable transitions is denoted by $\zeta ^a = \{ (y, \sigma, y') \in \zeta : \sigma \in \Sigma _o ^a \}$.
For each transition ${tr}=(y, \sigma, y')\in \zeta ^a$, the attacker can change the event $\sigma$ to any string in a language $A^{M}_{\sigma} \subseteq \Sigma ^*$, in a way similar to $A_{\sigma}$, but for transitions in $M$.
Note that, since the attacker observes $\Sigma _o$, it knows which state $y \in Y$ it is in. Hence, when it observes an event, it knows which transition the event corresponds to. Therefore, ALTER can be implemented based on $M$.


We ``embed'' $M$ into $G$ so that the attacks can be implemented based on $M$ as follows. Take the synchronous product \cite{cassandras2021introduction, wonham2019supervisory} of $G$ and $M$:
\begin{align*}
	\hat{G} & =( \hat{Q}, \Sigma , \hat{\delta} , \hat{q}_o ) = G \| M = (Q \times Y, \Sigma, \delta \times \zeta, (q_o, y_o)) .
\end{align*}
For a transition  $\hat{tr} =(\hat{q}, \sigma, \hat{q}') =((q, y), \sigma, (q' ,y')) \in \hat{G}$ with $\sigma \in \Sigma _o ^a$, the corresponding attack language is given by ${A}_{\hat{tr}} = A^{M}_{(y, \sigma, y')}$. Note that the automaton $M$ enables dynamic attacks so that the attack languages can now be different even for two transitions under the same event $\sigma$. The attack languages are still specified by $A_{\hat{tr}}$.
Since $P(L(G)) \subseteq L(M)$ implies $L(\hat{G}) = L(G)$, in the rest of the paper, we assume that, without loss of generality, $G$ is already embedded with some $M$. If not, we can always take $\hat{G} = G \| M$, call $\hat{G}$ the new $G$, and we work on the new $G$.

Let us illustrate the above discussions using the following example.

\begin{example}
	
	Let us consider the discrete event system $G$ in Fig. \ref{fig:GG} with states $Q= \{1, 2, 3, 4 \}$ and events $\Sigma = \{b, c, d, \}$.\footnote{We use software IDES developed in Karen Rudie's lab at Queens University (https://github.com/krudie/IDES) to create automata and perform operations on them.} In the figures, symbol $\rightarrow$ denotes the initial state and double circles denote marked states.

	\begin{figure}[hbt!]
		\centering
		\includegraphics[scale=0.6]{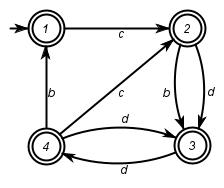}
		\caption{Discrete event system $G$ of Example 1.  }
		\label{fig:GG}
	\end{figure}

	Assume that event $b$ is unobservable and event $d$ is attackable, that is, $\Sigma_o = \{c, d \}$ and $\Sigma_o^a = \{ d \}$. The attacks are dynamic and are implemented by the automaton $M$ in Fig. \ref{fig:MM} as follows.
	In $M$, there are two attackable transitions: $\zeta ^a = \{ tr_1, tr_2 \}$, where $tr_1 = (II, d, III), tr_2=(III, d, III)$.
	Let $A^{M}_{tr_1} = \{ cd \}$ and $A^{M}_{tr_2} = \{ dc \}$. Note that, the attacker will insert $c$ before $d$ at state $II$ and insert $c$ after $d$ at state $III$.
	The corresponding automata $F^{M}_{tr}$ are shown in Fig. \ref{fig:FF2} and Fig, \ref{fig:FF1}, respectively.

	\begin{figure}[hbt!]
		\centering
		\includegraphics[scale=0.6]{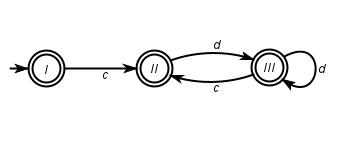}
		\caption{The automaton $M$ implementing the dynamic attack.  }
		\label{fig:MM}
	\end{figure}

	\begin{figure}[hbt!]
		\centering
		\includegraphics[scale=0.6]{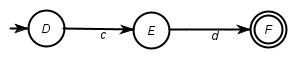}
		\caption{Automaton $F^{M}_{tr_1}$ of attack language $A^{M}_{tr_1}$ for transition $tr_1=(II, d, III) \in \zeta ^a$.  }
		\label{fig:FF2}
	\end{figure}

	\begin{figure}[hbt!]
		\centering
		\includegraphics[scale=0.6]{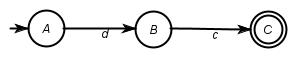}
		\caption{Automaton $F^{M}_{tr_2}$ of attack language $A^{M}_{tr_2}$ for transition $tr_2=(III, d, III) \in \zeta ^a$.  }
		\label{fig:FF1}
	\end{figure}

	The synchronous product $\hat{G} = G \| M$ is shown in Fig. \ref{fig:GM}.
	
	\begin{figure}[hbt!]
		\centering
		\includegraphics[scale=0.6]{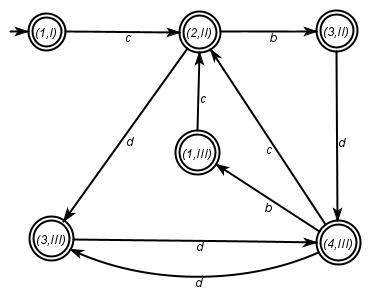}
		\caption{The synchronous product $\hat{G} = G \| M$ of Example 1.  }
		\label{fig:GM}
	\end{figure}

	There are four attackable transitions in $\hat{G}$: $\hat{\delta} ^a = \{ \hat{tr}_1, \hat{tr}_2, \hat{tr}_3, \hat{tr}_4 \}$, where $\hat{tr}_1 = ((2,II), d, (3,III))$, $\hat{tr}_2 = ((3,II), d, (4,III))$, $\hat{tr}_3 = ((3,III), d, (4,III))$, and $\hat{tr}_4 = ((4,III), d, (3,III))$.
	We have ${A}_{\hat{tr}_1} = {A}_{\hat{tr}_2} = A^{M}_{tr_1} = \{ cd \}$ and ${A}_{\hat{tr}_3} = {A}_{\hat{tr}_4} = A^{M}_{tr_2} = \{ dc \}$. The extended automaton $\hat{G}^a$ obtained after replacing all transitions subject to attacks in $\hat{G}$ is shown in Fig. \ref{fig:GMa}. Note that new states H,G,I,J are needed since they are not in $G$.
	
	\begin{figure}[hbt!]
		\centering
		\includegraphics[scale=0.6]{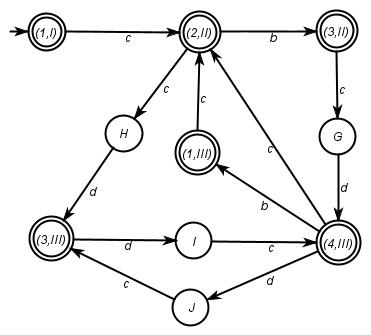}
		\caption{The extended automaton $\hat{G}^a$ of Example 1.  }
		\label{fig:GMa}
	\end{figure}
		
\end{example}

To model actuator attacks, we assume that the disablements/enablements of some controllable events in $\Sigma _c^a \subseteq \Sigma _c$ can be altered by attackers in the sense that an attacker can enable an event that is disabled by the supervisor or disable an event that is enabled by the supervisor.
In other words, for a given control (pattern) $\gamma$ (the set of enabled events issued by the supervisor such that $\Sigma_{uc} \subseteq \gamma \subseteq \Sigma$), an attacker can add some events in $\Sigma _c^a$ to $\gamma$ or remove some events in $\Sigma _c^a$ from $\gamma$. Hence, the set of possible controls after attacks is given by
\begin{align*}
	\Delta(\gamma) = \{ \gamma^a \in 2^{\Sigma} : (\exists \gamma ', \gamma ''
	\subseteq \Sigma _c^a ) \gamma^a = (\gamma\setminus\gamma ')\cup \gamma ''\} .
\end{align*}

Under joint sensor and actuator attacks, the supervised system behaves as follows. After the occurrence of $s\in L (G)$, a supervisor $\cS$ observes a string $w \in \Phi ^a (s)$. Based on $w$, $\cS$ enables a set of events, denoted by $\cS(w)$. Hence, $\cS$ is a mapping
$$
\cS :\Phi ^a ( L (G)) \rightarrow 2^{\Sigma}.
$$

The control command $\cS(w)$ issued by the supervisor after observing $w \in \Phi ^a ( L (G))$ may be altered by an attacker. We use $\cS^a (w)$ to denote the set of all possible control commands that may be received by the plant under actuator attacks, that is,
\begin{align*}
	\cS^a(w)=\Delta(\cS(w)).
\end{align*}

\begin{figure}[hbt!]
	\centering
	\includegraphics[scale=0.3]{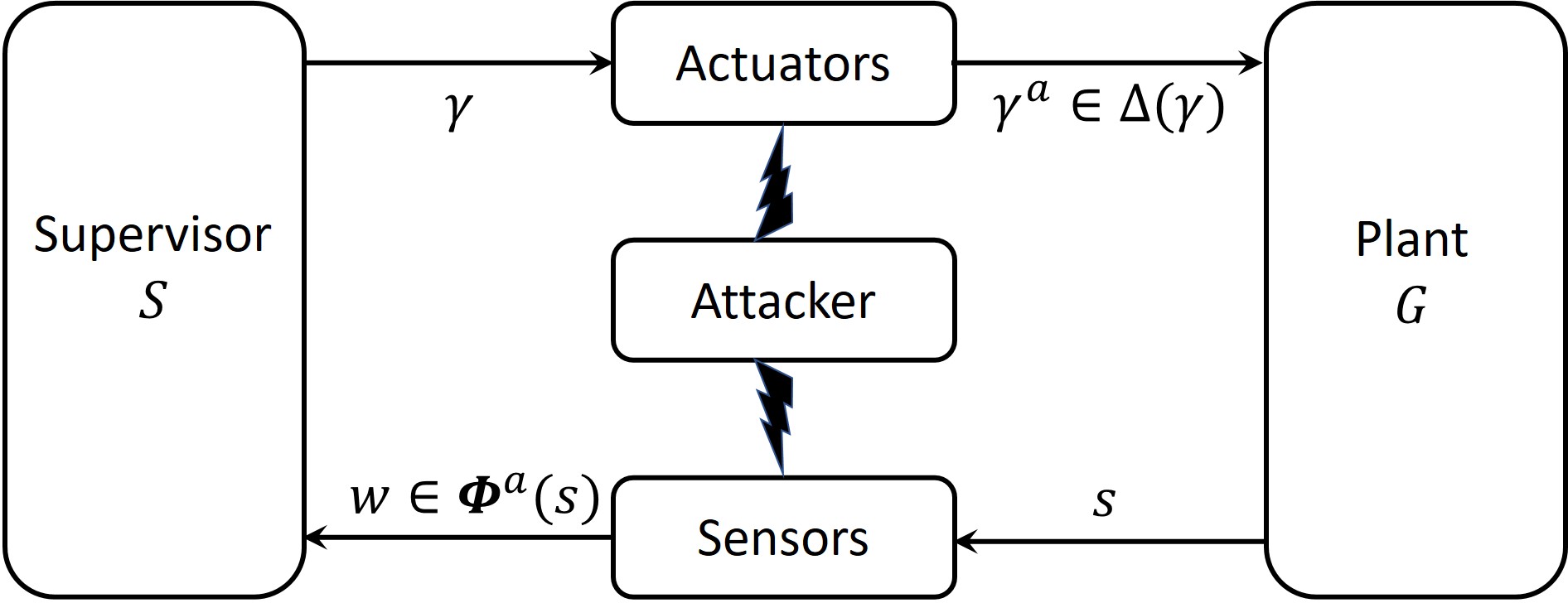}
	\caption{Supervised system under cyber attacks. }
	\label{fig:system}
\end{figure}

We denote the supervised system under cyber attacks as $\cS^a/G$, whose schematic is shown in Fig. \ref{fig:system}.
We refer the reader to \cite{zheng2023modeling} for the definition of $\cS^a/G$, but we will define local closed-loops as well as conjunctive closed-loop in the next section.

\section{Coordination Control under Cyber Attacks}

In this section, supervisory control under cyber attacks will be applied to modular systems.
Let $\Sigma _i, i=1,\ldots,n,$ be local alphabets (event sets) with $\Sigma= \cup_{i=1}^{n} \Sigma_i$ and $G_i  = (Q_i, \Sigma_i , \delta _i , q_{i,o})$ be local automata over $\Sigma_i$.
A modular DES is given by the synchronous product $G=G_1 \| \dots \| G_n$. Since in general a coordinator is needed to deal with global (indecomposable) specification languages $K\subseteq \Sigma^*$, our approach is based on so called coordination control
\cite{KMvScoordinationcontrolJDEDS15}.

\subsection{Coordination control of modular DES}

Without loss of generality, we assume that $n=2$. We consider a modular DES $G=G_1 \parallel G_2$. In coordination control, instead of computing a single supervisor $\cS$ for the global (monolithic) plant $G$, local supervisors $S_i$ are synthesized for components $G_i$, $i=1,2$.
Unfortunately, this requires the control specification language to be decomposable (or set of safe states in a form of Cartesian product), which is not always possible in general.
Therefore, coordination control has been proposed, where local supervisors $\cS_i$ are designed to control local plants extended by coordinator events that need to be communicated among local supervisors to jointly achieve the specification.
Concretely, $\cS_1$ will control $\tilde G_1=G_1 \| P_k(G_2)$ and $\cS_2$ will control $\tilde G_2=G_2 \| P_k(G_1)$, where $P_k: \Sigma^* \to \Sigma_k^*$ for some set of coordinator events $\Sigma_k \supseteq \Sigma_1 \cap \Sigma_2$ that depends on the specification, cf.  Section 4 or \cite{KMvScoordinationcontrolJDEDS15}.

It is well known that the computation of projected automaton $P_k(G_i)$ can be exponential in the worst case. However, it is also known that if the projection $P_k$ satisfies so called observer property with respect to the local plant language $L(G_i)$ \cite{wong1996hierarchical}, then the number of states of the minimal  generator for the projected local language, i. e.  $\| P_k(L(G_i)) \| $, is smaller or equal to $\| L(G_i) \| $.
Moreover, if the projection $P_k$ does not satisfy the observer property, then we can always enlarge $ \Sigma_k$ so that $P_i$ satisfies the observer property \cite{pena2008polynomial}.

To simplify the notation, we will denote $\tilde G_i$ by $G_i$ for $i=1,2$ in the sequel knowing that local plants are extended with coordinator parts whenever needed.

\subsection{Cyber attacks in coordination control}

Observable events and controllable events associated with local supervisor $\cS_i$ for $G_i, i=1,2$ are denoted by $\Sigma _{i,o} = \Sigma _i \cap \Sigma_o$ and $\Sigma _{i,c} = \Sigma _i \cap \Sigma_c$ respectively.
The set of attackable events in observation channel in $\Sigma _i$ is denoted by $\Sigma _{i,o} ^a =  \Sigma _{o} ^a \cap \Sigma _{i,o}=\Sigma _{o} ^a  \cap \Sigma_i$.
The set of attackable events in control channel in $\Sigma _i$ is denoted by $\Sigma _{i,c} ^a =  \Sigma _{c} ^a \cap \Sigma _{i,c}=\Sigma _{c} ^a \cap \Sigma _i$.
The projection from $\Sigma ^*$ to $\Sigma ^*_{i,o}$ is denoted by $P_{i,o} : \Sigma ^* \rightarrow \Sigma ^*_{i,o}$. Clearly, $P_{i,o} = P_{|i} \circ P_i$, where $\circ$ denotes composition and  $P_{|i}$ restriction of $P$ to $\Sigma_i^*$.

The observation mapping for $\cS _i$, denoted by $\Phi_i^a : L(G_i) \rightarrow 2^{\Sigma_{i,o} ^*}$ is given by, for $s \in L(G_i)$,
$$
\Phi^a_i (s) = P_{i,o} (\Theta_i ^a  (s)),
$$
where $\Theta_i ^a (s)$ describes the results of sensor attacks on local strings $s \in \Sigma_i^*$.
We can also specify $\Theta_i^a$ separately for $\Sigma_1$ and $\Sigma_2$, as long as they satisfy the following assumption, which ensures that the models for cyber attacks are consistent between $G$ and $G_i$.

\begin{assumption}
	The models for cyber attacks on $G$ and $G_i$ are consistent in the following sense:
	\begin{equation} \label{Assumption1}
		\begin{split}
			\Phi_i^a (P_i(s)) = \Phi_i ^a (s), i=1,2.
		\end{split}
	\end{equation}
\end{assumption}

In other words, we assume that cyber attacks on local events depend on local observations, which is a reasonable assumption.
Note that, without cyber attacks, Assumption 1 is equivalent to $P_i (P_i(s)) = P_i(s)$, which is obviously true.
Note further that, since static attacks are independent of observations, Assumption 1 is automatically satisfied.

After the occurrence of $s \in L(G)$, supervisor $\cS_i$ observes a string $w_i \in \Phi^a_i (s)$ and issues a control $\cS_i(w_i) \subseteq \Sigma _i$. Hence, $\cS_i$ is a mapping
$$
\cS_i :\Phi_i ^a (L (G)) \rightarrow 2^{\Sigma_i}.
$$
Since uncontrollable events cannot be disabled, we require that $\Sigma _{i,uc} \subseteq \cS_i(w_i)$.
The control $\cS_i(w_i)$ may be altered by an attacker similar to that of the centralized supervisor as
\begin{align*}
	\cS^a_i(w_i)=\Delta_i (\cS_i(w_i)),
\end{align*}
where $\Delta_i (.)$ is defined as follows.
For $\gamma_i$ such that $\Sigma _{i,c} \subseteq \gamma_i \subseteq \Sigma _i$,
$$
\Delta_i (\gamma_i) = \{ \gamma_i^a \in 2^{\Sigma_i} : (\exists \gamma_i ', \gamma_i ''
\subseteq \Sigma _{i,c}^a ) \gamma_i^a = (\gamma_i \setminus \gamma_i ') \cup \gamma_i ''\}.
$$

Because both observation and control are nondeterministic (multi-valued), the language generated by the supervised system $L (\cS_i^a/G_i)$ is also nondeterministic, and hence may not be unique.
We consider the upper bound of all possible languages that can be generated by the supervised system $\cS_i^a/G_i$. We call the upper bound {\em large (allowed) language} and denote it by $L _a (\cS_i^a/G_i)$. $L _a (\cS_i^a/G_i)$ is defined recursively as follows.
\begin{enumerate}
	\item
	The empty string belongs to $L_a(\cS_i^a/G_i)$, that is,
	$$
	\varepsilon \in L_a(\cS_i^a/G_i).
	$$
	\item
	If $s$ belongs to $L_a(\cS_i^a/G_i)$, then for any $\sigma \in 	\Sigma_i$, $s \sigma$ belongs to $L_a(\cS_i^a/G_i)$ if and only if $s \sigma$ is physically possible in $L(G_i)$ and $\sigma$ is uncontrollable or enabled by $\cS_i^a$ in some situations.
	Formally, for any $s \in L_a(\cS_i^a/G_i)$ and $\sigma \in \Sigma_i$,
	\begin{align*}
		 s \sigma\in L_a(\cS_i^a/G_i) 	\Leftrightarrow  s \sigma \in L (G_i) \wedge (\sigma \in \Sigma _{uc}
		 \vee (\exists w \in \Phi_i^a (s))(\exists \gamma_i^a \in \cS_i^a(w)) \sigma \in \gamma_i^a) .
	\end{align*}
\end{enumerate}

The large language $L _a (\cS_i^a/G_i)$ can be used to ensure safety, that is, all possible strings generated by the supervised system are contained in a safe/legal specification language under all possible sensor and actuator attacks.
Note that the large language may contain strings that are not possible in the supervised system.

The necessary and sufficient conditions for the existence of a supervisor $\cS_i$ such that $L_a(\cS_i^a/G_i) = K_i$ can be expressed in terms of CA-controllability and CA-observability defined as follows.

\begin{definition}
	A closed language $K_i \subseteq L(G_i)$ is {\em CA-controllable} with respect to $L(G_i)$, $\Sigma _{uc} \cap \Sigma_i$, and $\Sigma ^a_c \cap \Sigma_i$ if
	\begin{equation} \label{C}
		\begin{split}
			K_i (( \Sigma_{uc} \cup \Sigma_c^a ) \cap \Sigma_i) \cap L(G_i) \subseteq K_i.
		\end{split}
	\end{equation}
\end{definition}

In other words, $K_i$ is {\em CA-controllable} with respect to $L(G_i)$, $\Sigma _{uc} \cap \Sigma_i$, and $\Sigma ^a_c \cap \Sigma_i$ if and only if $K_i$ is controllable with respect to $L(G_i)$ and $(\Sigma _{uc} \cap \Sigma_i) \cup (\Sigma ^a_c \cap \Sigma_i)$.
Intuitively, this is because attackers can enable events in $\Sigma ^a_c$ and hence make them uncontrollable as far as the large language is concerned.

\begin{definition}
	A closed language $K_i \subseteq L(G_i)$ is {\em CA-observable} with respect to $L(G_i)$, $\Sigma _{o} \cap \Sigma_i$, and $\Phi _i^a$ if
	\begin{equation} \label{Equation7}
		\begin{split}
			 (\forall s \in \Sigma_i ^*)(\forall \sigma \in \Sigma_i) (s \sigma \in K_i  \Rightarrow & (\exists w \in \Phi_i ^a (s))(\forall s' \in (\Phi_i ^a)^{-1} (w)) (s' \in K_i \wedge s' \sigma \in L(G_i) \Rightarrow s' \sigma \in K_i)),
		\end{split}
	\end{equation}
	where $(\Phi_i ^a)^{-1}$ is the inverse mapping of $\Phi_i ^a$: for $w \in \Phi_i ^a (L(G_i))$, $(\Phi_i ^a)^{-1} (w) = \{ s' \in L(G_i) : w \in \Phi_i ^a (s') \}$.
\end{definition}

If there are no sensor attacks, that is, $\Phi_i ^a (s) = P_{i,o} (s)$ for all $s \in L(G_i)$, then CA-observability reduces to
\begin{align*}
	& (\forall s \in \Sigma_i ^*)(\forall \sigma \in \Sigma_i) (s \sigma \in K_i  \Rightarrow (\forall s' \in P_{i,o}^{-1} (P_{i,o}(s)))
	(s' \in K_i \wedge s' \sigma \in L(G_i) \Rightarrow s' \sigma \in K_i)),
\end{align*}
which is the conventional observability \cite{lin1988observability}.

While CA-controllability can be ``converted'' to conventional controllability by letting attackable events be uncontrollable, CA-observability can be ``converted'' to conventional observability. This fact shows the difficulties of handling observations under cyber attacks. It also means a new theory needs to be developed for CA-observability and coordination control of DES under cyber attacks as described below.

Using CA-controllability and CA-observability, we have the
following theorem.

\begin{theorem} \label{theoreme1}	
	Consider discrete event system $G_i, i=1,2$ under joint sensor and actuator attacks. For a nonempty closed language $K_i \subseteq L(G_i)$, there exists a local supervisor $\cS_i$ such that $L_a(\cS_i^a/G_i) = K_i$
	if and only if $K_i$ is CA-controllable (with respect to $L(G_i)$, $\Sigma _{uc} \cap \Sigma_i$, and $\Sigma ^a_c \cap \Sigma_i$) and CA-observable (with respect to $L(G_i)$, $\Sigma _{o} \cap \Sigma_i$, and $\Phi _i^a$)
	
\end{theorem}
\noindent {\em Proof:}
The proof is similar to that of Theorem 3 in \cite{zheng2023modeling}.
\hfill \bull

\subsection{Relation between local and global large languages}

We consider supervisors $\cS_1$ and $\cS_2$ working together in conjunction (denoted by $\wedge$) in the sense that a common event $\sigma \in \Sigma _1 \cap \Sigma _2$ is enabled if and only if it is enabled by both $\cS_1$ and $\cS_2$.
On the other hand, events in $\Sigma _1 \setminus \Sigma _2$ ($\Sigma _2 \setminus \Sigma _1$, respectively) are enabled if and only if they are enabled by $\cS_1$ ($\cS_2$, respectively).

The supervised system with coordination supervisors under cyber attacks is denoted by $(\cS_1^a \wedge \cS_2^a)/G$. The large language generated by $(\cS_1^a \wedge \cS_2^a)/G$ is denoted by $L_a((\cS_1^a \wedge \cS_2^a)/G)$. It is defined recursively as follows.
\begin{enumerate}
	\item
	The empty string belongs to $L_a((\cS_1^a \wedge \cS_2^a)/G)$, that is,
	$$
	\varepsilon \in L_a((\cS_1^a \wedge \cS_2^a)/G).
	$$
	\item
	If $s$ belongs to $L_a((\cS_1^a \wedge \cS_2^a)/G)$, then for any $\sigma \in \Sigma$, $s \sigma$ belongs to $L_a((\cS_1^a \wedge \cS_2^a)/G)$
	if and only if $s \sigma$ is physically possible in $L(G)$ and $\sigma$ is uncontrollable or enabled by $\cS_1^a \wedge \cS_2^a$ in some situations.
	Formally, for any $s \in L_a((\cS_1^a \wedge \cS_2^a)/G)$ and $\sigma \in \Sigma$,
	\begin{align*}
		& s \sigma\in L_a((\cS_1^a \wedge \cS_2^a)/G) \\
		\Leftrightarrow
		& s \sigma \in L (G) \wedge (\sigma \in \Sigma _{uc}
		 \vee (\exists w_1 \in \Phi_1 ^a (s))(\exists w_2 \in \Phi_2 ^a (s)) \\
		& (\exists \gamma_1^a \in \cS_1^a(w_1)) (\exists \gamma_2^a \in \cS_2^a(w_2))
		 \sigma \in (\gamma^a_1 \cup (\Sigma \setminus \Sigma_1)) \cap (\gamma^a_2 \cup (\Sigma \setminus \Sigma_2))) .
	\end{align*}
\end{enumerate}

Let us first prove the following relations between $L_a((\cS_1^a \wedge \cS_2^a)/G)$ and $L_a(\cS_i^a/G_i)$.

\begin{theorem} \label{LGS12}
The large language generated by $(\cS_1^a \wedge \cS_2^a)/G$ is given by
\begin{align*}
	& L_a((\cS_1^a \wedge \cS_2^a)/G) = L_a(\cS_1^a/G_1) \parallel  L_a(\cS_2^a/G_2) .
\end{align*}

\end{theorem}

\noindent {\em Proof:} We prove, for all $s \in L(G)$,
\begin{align*}
	 s \in L_a((\cS_1^a \wedge \cS_2^a)/G)
	\Leftrightarrow
	 s \in L_a(\cS_1^a/G_1) \parallel  L_a(\cS_2^a/G_2)  .
\end{align*}
by induction on the length $|s|$.

{\em Base:} By definition, $\varepsilon \in L_a((\cS_1^a \wedge \cS_2^a)/G)$, $\varepsilon \in L_a(\cS_1^a/G_1)$, and $\varepsilon \in L_a(\cS_2^a/G_2)$. Therefore, for
$|s| = 0$, that is, $s=\varepsilon$, we have
\begin{align*}
	 s \in L_a((\cS_1^a \wedge \cS_2^a)/G)
	\Leftrightarrow
	 s \in L_a(\cS_1^a/G_1) \parallel  L_a(\cS_2^a/G_2)  .
\end{align*}
{\em Induction Hypothesis:} Assume that for all $s \in \Sigma^*$,
$|s| \leq m$,
\begin{align*}
	 s \in L_a((\cS_1^a \wedge \cS_2^a)/G)
	\Leftrightarrow
	 s \in L_a(\cS_1^a/G_1) \parallel  L_a(\cS_2^a/G_2)  .
\end{align*}
{\em Induction Step:} We show that for all $s \in \Sigma^*$,
$\sigma \in \Sigma$, $|s\sigma| = m+1$,
\begin{align*}
	 s \sigma \in L_a((\cS_1^a \wedge \cS_2^a)/G)
	\Leftrightarrow
	 s \sigma \in L_a(\cS_1^a/G_1) \parallel  L_a(\cS_2^a/G_2) ,
\end{align*}
by considering three cases.

\noindent
{\em Case 1}: $\sigma \in \Sigma _1 \setminus \Sigma_2$. In this case, we have
\begin{align*}
	& s \sigma\in L_a((\cS_1^a \wedge \cS_2^a)/G) \\
	\Leftrightarrow
	& s \in L_a((\cS_1^a \wedge \cS_2^a)/G) \wedge s \sigma \in L (G) \wedge (\sigma \in \Sigma _{uc}  \vee (\exists w_1 \in \Phi_1 ^a (s)) (\exists w_2 \in \Phi_2 ^a (s)) \\
    & (\exists \gamma_1^a \in \cS_1^a(w_1)) (\exists \gamma_2^a \in \cS_2^a(w_2)) \sigma \in (\gamma^a_1 \cup (\Sigma \setminus \Sigma_1)) \cap (\gamma^a_2 \cup (\Sigma \setminus \Sigma_2))) \\
	& (\mbox{by the definition of } L_a((\cS_1^a \wedge \cS_2^a)/G)) \\
	\Leftrightarrow
	& s \in L_a((\cS_1^a \wedge \cS_2^a)/G) \wedge s \sigma \in L (G) \wedge (\sigma \in \Sigma _{uc} \\
	& \vee (\exists w_1 \in \Phi_1 ^a (s)) (\exists w_2 \in \Phi_2 ^a (s))(\exists \gamma_1^a \in \cS_1^a(w_1))
	 (\exists \gamma_2^a \in \cS_2^a(w_2)) \sigma \in \gamma^a_1) \\
	& (\mbox{because } \sigma \in \Sigma _1 \setminus \Sigma_2 \Rightarrow \sigma \not\in \Sigma \setminus \Sigma_1 \wedge \sigma \in \Sigma \setminus \Sigma_2) \\
	\Leftrightarrow
	& s \in L_a((\cS_1^a \wedge \cS_2^a)/G) \wedge s \sigma \in L (G) \wedge (\sigma \in \Sigma _{uc} \vee (\exists w_1 \in \Phi_1 ^a (s)) (\exists \gamma_1^a \in \cS_1^a(w_1)) \sigma \in \gamma^a_1 ) .
\end{align*}
On the other hand,
\begin{align*}
	& s \sigma \in L_a(\cS_1^a/G_1) \parallel  L_a(\cS_2^a/G_2) \\
	\Leftrightarrow
	& s \in L_a(\cS_1^a/G_1) \parallel  L_a(\cS_2^a/G_2)
	 \wedge s \sigma \in L_a(\cS_1^a/G_1) \parallel  L_a(\cS_2^a/G_2) \\
	& (\mbox{because $L_a(\cS_1^a/G_1) \parallel  L_a(\cS_2^a/G_2)$ is closed } ) \\
	\Leftrightarrow
	& s \in L_a((\cS_1^a \wedge \cS_2^a)/G)
	 \wedge s \sigma \in L_a(\cS_1^a/G_1) \parallel  L_a(\cS_2^a/G_2) \\
	& (\mbox{by Induction Hypothesis}) \\
	\Leftrightarrow
	& s \in L_a((\cS_1^a \wedge \cS_2^a)/G) \wedge s \sigma \in L (G)
	 \wedge s \sigma \in L_a(\cS_1^a/G_1) \parallel  L_a(\cS_2^a/G_2) \\
	& (\mbox{because } L_a(\cS_1^a/G_1) \parallel  L_a(\cS_2^a/G_2)
	 \subseteq L(G_1) \parallel  L(G_2) = L(G) ) \\
	\Leftrightarrow
	& s \in L_a((\cS_1^a \wedge \cS_2^a)/G) \wedge s \sigma \in L (G) \wedge P_1(s \sigma) \in L_a(\cS_1^a/G_1)  \wedge P_2(s \sigma) \in L_a(\cS_2^a/G_2) \\
	& (\mbox{by the definition of } \parallel  ) \\
	\Leftrightarrow
	& s \in L_a((\cS_1^a \wedge \cS_2^a)/G) \wedge s \sigma \in L (G)  \wedge P_1(s) \sigma \in L_a(\cS_1^a/G_1)  \wedge P_2(s) \in L_a(\cS_2^a/G_2) \\
	& (\mbox{because } \sigma \in \Sigma _1 \setminus \Sigma_2 ) \\
	\Leftrightarrow
	& s \in L_a((\cS_1^a \wedge \cS_2^a)/G) \wedge s \sigma \in L (G)
	 \wedge P_1(s) \sigma\in L_a(\cS_1^a/G_1) \\
	& (\mbox{by Induction Hypothesis, }
	 s \in L_a((\cS_1^a \wedge \cS_2^a)/G) \Rightarrow P_2(s) \in L_a(\cS_2^a/G_2) ) \\
	\Leftrightarrow
	& s \in L_a((\cS_1^a \wedge \cS_2^a)/G) \wedge s \sigma \in L (G)  \wedge P_1(s) \in L_a(\cS_1^a/G_1) \wedge P_1(s) \sigma \in L (G_1) \\
    & \wedge (\sigma \in \Sigma _{uc} \vee (\exists w_1 \in \Phi_1^a (P_1(s)))(\exists \gamma_1^a \in \cS_1^a(w_1)) \sigma \in \gamma_1^a) \\
	& (\mbox{by the definition of } L_a(\cS_1^a/G_1)) \\
	\Leftrightarrow
	& s \in L_a((\cS_1^a \wedge \cS_2^a)/G) \wedge s \sigma \in L (G) \wedge (\sigma \in \Sigma _{uc}
	 \vee (\exists w_1 \in \Phi_1^a (P_1(s)))(\exists \gamma_1^a \in \cS_1^a(w_1)) \sigma \in \gamma_1^a) \\
	& (\mbox{because } s \in L_a((\cS_1^a \wedge \cS_2^a)/G) \wedge s \sigma \in L (G)
	 \Rightarrow P_1(s) \in L_a(\cS_1^a/G_1) \wedge P_1(s) \sigma \in L (G_1) )\\
	\Leftrightarrow
	& s \in L_a((\cS_1^a \wedge \cS_2^a)/G) \wedge s \sigma \in L (G)
	 \wedge (\sigma \in \Sigma _{uc} \vee (\exists w_1 \in \Phi_1 ^a (s)) (\exists \gamma_1^a \in \cS_1^a(w_1)) \sigma \in \gamma^a_1 ) \\
	& (\mbox{by Assumption 1}).
\end{align*}
Therefore,
\begin{align*}
	& s \sigma\in L_a((\cS_1^a \wedge \cS_2^a)/G) \\
	\Leftrightarrow
	& s \in L_a((\cS_1^a \wedge \cS_2^a)/G) \wedge s \sigma \in L (G)
	 \wedge (\sigma \in \Sigma _{uc} \vee (\exists w_1 \in \Phi_1 ^a (s)) (\exists \gamma_1^a \in \cS_1^a(w_1)) \sigma \in \gamma^a_1 ) \\
	\Leftrightarrow
	& s \sigma \in L_a(\cS_1^a/G_1) \parallel  L_a(\cS_2^a/G_2) .
\end{align*}

\noindent
{\em Case 2}: $\sigma \in \Sigma _2 \setminus \Sigma_1$. This case is similar to Case 1 with subscripts 1 and 2 interchanged.\\

\noindent
{\em Case 3}: $\sigma \in \Sigma _1 \cap \Sigma_2$. In this case, we have
\begin{align*}
	& s \sigma\in L_a((\cS_1^a \wedge \cS_2^a)/G) \\
	\Leftrightarrow
	& s \in L_a((\cS_1^a \wedge \cS_2^a)/G) \wedge s \sigma \in L (G) \wedge (\sigma \in \Sigma _{uc} \vee (\exists w_1 \in \Phi_1 ^a (s)) (\exists w_2 \in \Phi_2 ^a (s)) \\
	& (\exists \gamma_1^a \in \cS_1^a(w_1))(\exists \gamma_2^a \in \cS_2^a(w_2)) \sigma \in (\gamma^a_1 \cup (\Sigma \setminus \Sigma_1)) \cap (\gamma^a_2 \cup (\Sigma \setminus \Sigma_2))) \\
	& (\mbox{by the definition of } L_a((\cS_1^a \wedge \cS_2^a)/G)) \\
	\Leftrightarrow
	& s \in L_a((\cS_1^a \wedge \cS_2^a)/G) \wedge s \sigma \in L (G) \wedge (\sigma \in \Sigma _{uc} \\
	& \vee (\exists w_1 \in \Phi_1 ^a (s)) (\exists w_2 \in \Phi_2 ^a (s))(\exists \gamma_1^a \in \cS_1^a(w_1))
	 (\exists \gamma_2^a \in \cS_2^a(w_2)) \sigma \in \gamma^a_1 \cap \gamma^a_2 ) \\
	& (\mbox{because } \sigma \in \Sigma _1 \cap \Sigma_2 \Rightarrow \sigma \not\in \Sigma \setminus \Sigma_1 \wedge \sigma \not\in \Sigma \setminus \Sigma_2) \\
	\Leftrightarrow
	& s \in L_a((\cS_1^a \wedge \cS_2^a)/G) \wedge s \sigma \in L (G) \wedge (\sigma \in \Sigma _{uc} \\
	& \vee (\exists w_1 \in \Phi_1 ^a (s)) (\exists w_2 \in \Phi_2 ^a (s))(\exists \gamma_1^a \in \cS_1^a(w_1))
	 (\exists \gamma_2^a \in \cS_2^a(w_2)) \sigma \in \gamma^a_1 \wedge \sigma \in \gamma^a_2 ) \\
	\Leftrightarrow
	& s \in L_a((\cS_1^a \wedge \cS_2^a)/G) \wedge s \sigma \in L (G) \wedge (\sigma \in \Sigma _{uc} \\
	& \vee ((\exists w_1 \in \Phi_1 ^a (s)) (\exists \gamma_1^a \in \cS_1^a(w_1)) \sigma \in \gamma^a_1
	 \wedge (\exists w_2 \in \Phi_2 ^a (s)) (\exists \gamma_2^a \in \cS_2^a(w_2))  \sigma \in \gamma^a_2 )) \\
	\Leftrightarrow
	& s \in L_a((\cS_1^a \wedge \cS_2^a)/G) \wedge s \sigma \in L (G) \\
	& \wedge (\sigma \in \Sigma _{uc} \vee (\exists w_1 \in \Phi_1 ^a (s)) (\exists \gamma_1^a \in \cS_1^a(w_1)) \sigma \in \gamma^a_1 )  \wedge (\sigma \in \Sigma _{uc} \vee (\exists w_2 \in \Phi_2 ^a (s)) (\exists \gamma_2^a \in \cS_2^a(w_2)) \sigma \in \gamma^a_2 ) .
\end{align*}
On the other hand,
\begin{align*}
	& s \sigma \in L_a(\cS_1^a/G_1) \parallel  L_a(\cS_2^a/G_2) \\
	\Leftrightarrow
	& s \in L_a((\cS_1^a \wedge \cS_2^a)/G) \wedge s \sigma \in L (G)
	 \wedge P_1(s \sigma) \in L_a(\cS_1^a/G_1)  \wedge P_2(s \sigma) \in L_a(\cS_2^a/G_2) \\
	& (\mbox{see Case 1} ) \\
	\Leftrightarrow
	& s \in L_a((\cS_1^a \wedge \cS_2^a)/G) \wedge s \sigma \in L (G)
	 \wedge P_1(s) \sigma \in L_a(\cS_1^a/G_1)  \wedge P_2(s) \sigma \in L_a(\cS_2^a/G_2) \\
	& (\mbox{because } \sigma \in \Sigma _1 \cap \Sigma_2 ) \\
	\Leftrightarrow
	& s \in L_a((\cS_1^a \wedge \cS_2^a)/G) \wedge s \sigma \in L (G)  \wedge P_1(s) \in L_a(\cS_1^a/G_1) \wedge P_1(s) \sigma \in L (G_1) \\
    & \wedge (\sigma \in \Sigma _{uc}  \vee (\exists w_1 \in \Phi_1^a (P_1(s)))(\exists \gamma_1^a \in \cS_1^a(w_1)) \sigma \in \gamma_1^a)  \wedge P_2(s) \in L_a(\cS_2^a/G_2) \\
    & \wedge P_2(s) \sigma \in L (G_2) \wedge (\sigma \in \Sigma _{uc}  \vee (\exists w_2 \in \Phi_2^a (P_2(s)))(\exists \gamma_2^a \in \cS_2^a(w_2)) \sigma \in \gamma_2^a) \\
	& (\mbox{by the definition of } L_a(\cS_i^a/G_i)) \\
	\Leftrightarrow
	& s \in L_a((\cS_1^a \wedge \cS_2^a)/G) \wedge s \sigma \in L (G)  \wedge (\sigma \in \Sigma _{uc} \vee (\exists w_1 \in \Phi_1^a (P_1(s)))
	 (\exists \gamma_1^a \in \cS_1^a(w_1)) \sigma \in \gamma_1^a) \\
	& \wedge (\sigma \in \Sigma _{uc} \vee (\exists w_2 \in \Phi_2^a (P_2(s)))
	 (\exists \gamma_2^a \in \cS_2^a(w_2)) \sigma \in \gamma_2^a) \\
	& (\mbox{because } s \in L_a((\cS_1^a \wedge \cS_2^a)/G) \wedge s \sigma \in L (G) \Rightarrow P_i(s) \in L_a(\cS_i^a/G_i) \wedge P_i(s) \sigma \in L (G_i) )\\
	\Leftrightarrow
	& s \in L_a((\cS_1^a \wedge \cS_2^a)/G) \wedge s \sigma \in L (G)  \wedge (\sigma \in \Sigma _{uc} \vee (\exists w_1 \in \Phi_1 ^a (s)) (\exists \gamma_1^a \in \cS_1^a(w_1)) \sigma \in \gamma^a_1 ) \\
	& \wedge (\sigma \in \Sigma _{uc} \vee (\exists w_2 \in \Phi_2 ^a (s)) (\exists \gamma_2^a \in \cS_2^a(w_2)) \sigma \in \gamma^a_2 ) \\
	& (\mbox{by Assumption 1})  .
\end{align*}
Therefore,
\begin{align*}
	& s \sigma\in L_a((\cS_1^a \wedge \cS_2^a)/G) \\
	\Leftrightarrow
	& s \in L_a((\cS_1^a \wedge \cS_2^a)/G) \wedge s \sigma \in L (G)
	 \wedge (\sigma \in \Sigma _{uc} \vee (\exists w_1 \in \Phi_1 ^a (s)) (\exists \gamma_1^a \in \cS_1^a(w_1)) \sigma \in \gamma^a_1 ) \\
	& \wedge (\sigma \in \Sigma _{uc} \vee (\exists w_2 \in \Phi_2 ^a (s)) (\exists \gamma_2^a \in \cS_2^a(w_2)) \sigma \in \gamma^a_2 ) \\
	\Leftrightarrow
	& s \sigma \in L_a(\cS_1^a/G_1) \parallel  L_a(\cS_2^a/G_2) .
\end{align*}
\hfill \bull

Theorem \ref{LGS12} tells us how to calculate the global large language from the local large languages if Assumption 1 is satisfied. Based on Theorem \ref{LGS12}, we investigate how to design coordination control in the next section.

\section{Design of Coordination Control under Cyber Attacks}

In this section, we investigate how to design coordination supervisors to achieve some control tasks. Control tasks are often specified by an admissible/legal language $K$ such that $K \subseteq L(G)$ \cite{ramadge1987supervisory, lin1988observability, cassandras2021introduction, wonham2019supervisory}.

In conventional supervisory control, a supervisor is designed such that $L(\cS/G) \subseteq K$. For supervisory control under cyber attacks, since the language generated by the supervised system is nondeterministic, we need to design a supervisor
$\cS :\Phi ^a ( L (G)) \rightarrow 2^{\Sigma}$  taking into account the sensor attacks such that, for $\cS^a$ (after actuator attacks), we still have $L_a(\cS^a/G) \subseteq K$.
The reason to use the large language in $L_a(\cS^a/G) \subseteq K$ is to ensure that no matter what language is actually generated by $\cS^a/G$, it is always contained in $K$.

In coordination control, we would like to design supervisors $\cS_i, i=1,2$, such that $L_a((\cS_1^a \wedge \cS_2^a)/G) \subseteq K$. To this end, we need to find necessary and sufficient conditions for the existence of supervisors $\cS_i, i=1,2$ such that $L_a((\cS_1^a \wedge \cS_2^a)/G)=K$ for a given language $K \subseteq L(G)$.

By Theorem \ref{LGS12}, we have $L_a((\cS_1^a \wedge \cS_2^a)/G) = L_a(\cS_1^a/G_1)$ $\parallel L_a(\cS_2^a/G_2)$.
Hence, in order to achieve the goal of $L_a((\cS_1^a \wedge \cS_2^a)/G)=K$, it is required that $K$ is conditionally decomposable with respect $\Sigma _1$ and $\Sigma _2$ \cite{komenda2012supervisory}, that is,
\begin{equation} \label{decomposable}
	\begin{split}
		K = P_1(K) \parallel  P_2 (K).
	\end{split}
\end{equation}
The above equation says that the synchronous product of the projections of $K$ to $\Sigma _1$ and $\Sigma _2$ is equal to $K$.

Note that the definition of conditional decomposability in \cite{komenda2012supervisory} is expressed as
\begin{equation} \label{decomposable2}
    \begin{split}
        K = P_{1+k}(K) \parallel  P_{2+k} (K),
    \end{split}
\end{equation}
where $P_{i+k} : \Sigma^* \to (\Sigma_i \cup \Sigma_k)^*$ is the projection from $\Sigma^*$ to $(\Sigma_i \cup \Sigma_k)^*$ (with $\Sigma_k$ being the coordinator alphabet).
Since we denote $\Sigma_i \cup \Sigma_k$ by $\Sigma_i$ to simplify the notation, Equation (\ref{decomposable2}) is simplified as Equation (\ref{decomposable}).
Recall that since $\Sigma_1 \cap \Sigma_2\subseteq \Sigma_i$, for $i=1,2$, the  natural projection distributes with the synchronous product, that is, for $i=1,2$ and any $L_i\subseteq \Sigma_i^*$
\begin{equation}
\label{projection_distributes}
P_i(L_1 \| L_2)=P_i(L_1) \| P_i(L_2).
\end{equation}

Note also that if $K$ is not decomposable with respect to original local alphabets $\Sigma_1$ and $\Sigma_2$, that is, $ P_{1}(K) \parallel  P_{2} (K)\not\subseteq K$,  then we can always enlarge $\Sigma _1$ and $\Sigma _2$ with coordinator events to make $K$ conditionally decomposable.
A polynomial time algorithm to do so is given in \cite{KvSMconddec2012}. Hence, in the rest of the paper, we assume that $K$ is conditionally decomposable.

\begin{theorem} \label{NSC}
There exist local supervisors $\cS_1$ and $\cS_2$ such that
\begin{align*}
	& L_a((\cS_1^a \wedge \cS_2^a)/G)=K
	 \wedge L_a(\cS_i^a/G_i) \subseteq P_i (K) , i=1,2
\end{align*}
if and only if (1) $K$ is conditionally decomposable; (2) $P_i(K)$ is CA-controllable (with respect to $L(G_i)$, $\Sigma _{uc} \cap \Sigma_i$, and $\Sigma ^a_c \cap \Sigma_i$); and (3) $P_i(K)$ is CA-observable (with respect to $L(G_i)$, $\Sigma _{o} \cap \Sigma_i$, and $\Phi _i^a$), $i=1,2$.
	
\end{theorem}
\noindent {\em Proof:} 
(IF) Assume that (1) $K$ is conditionally decomposable; (2) $P_i(K)$ is CA-controllable (with respect to $L(G_i)$, $\Sigma _{uc} \cap \Sigma_i$, and $\Sigma ^a_c \cap \Sigma_i$); and (3) $P_i(K)$ is CA-observable (with respect to $L(G_i)$, $\Sigma _{o} \cap \Sigma_i$, and $\Phi _i^a$), $i=1,2$.
By Theorem \ref{theoreme1}, there exist local supervisors $\cS_i$ such that $L_a(\cS_i^a/G_i) = P_i(K)$. Clearly,
$$
L_a(\cS_i^a/G_i) = P_i(K) \Rightarrow L_a(\cS_i^a/G_i) \subseteq P_i (K).
$$
On the other hand, by Theorem \ref{LGS12},
\begin{align*}
	& L_a((\cS_1^a \wedge \cS_2^a)/G) \\
	= & L_a(\cS_1^a/G_1) \parallel  L_a(\cS_2^a/G_2)  \\
	= & P_1(K) \parallel  P_2(K)  \\
	= & K  \mbox{\ \ \ \ } (\mbox{because $K$ is conditionally decomposable})  .
\end{align*}
Therefore,
\begin{align*}
	& L_a((\cS_1^a \wedge \cS_2^a)/G)=K \wedge L_a(\cS_i^a/G_i) \subseteq P_i (K) , i=1,2 .
\end{align*}
(ONLY IF) Assume that there exist supervisors $\cS_1$ and $\cS_2$ such that
\begin{align*}
	& L_a((\cS_1^a \wedge \cS_2^a)/G)=K \wedge L_a(\cS_i^a/G_i) \subseteq P_i (K) , i=1,2 .
\end{align*}
Then, we have, for $i=1,2$,
\begin{align*}
	& L_a((\cS_1^a \wedge \cS_2^a)/G)=K \\
	\Rightarrow
	& K = L_a(\cS_1^a/G_1) \parallel  L_a(\cS_2^a/G_2)
	 \mbox{\ \ \ \ } (\mbox{by Theorem \ref{LGS12}}) \\
	\Rightarrow
	& P_i(K) = P_i(L_a(\cS_1^a/G_1) \parallel  L_a(\cS_2^a/G_2) ) \\
	\Rightarrow
	& P_i(K) = P_i(L_a(\cS_1^a/G_1)) \parallel  P_i (L_a(\cS_2^a/G_2) )
	 \mbox{\ \ \ \ } (\mbox{by Equation (\ref{projection_distributes})}) \\
	\Rightarrow
	& P_i(K) \subseteq P_i(L_a(\cS_i^a/G_i)) = L_a(\cS_i^a/G_i) .
\end{align*}
Since $L_a(\cS_i^a/G_i) \subseteq P_i (K)$ by the assumption, we have
$$
L_a(\cS_i^a/G_i) = P_i(K).
$$
By Theorem \ref{theoreme1}, $P_i(K)$ is CA-controllable (with respect to $L(G_i)$, $\Sigma _{uc} \cap \Sigma_i$, and $\Sigma ^a_c \cap \Sigma_i$) and CA-observable (with respect to $L(G_i)$, $\Sigma _{o} \cap \Sigma_i$, and $\Phi _i^a$).

By Theorem \ref{LGS12},
\begin{align*}
	K & = L_a((\cS_1^a \wedge \cS_2^a)/G) \\
	& = L_a(\cS_1^a/G_1) \parallel  L_a(\cS_2^a/G_2)  \\
	& = P_1(K) \parallel  P_2(K),
\end{align*}
that is, $K$ is conditionally decomposable.
\hfill \bull

Theorem \ref{NSC} gives the necessary and sufficient conditions for the existence of coordination control. If the conditions are satisfied, that is, if $K$ is conditionally decomposable, and $P_i(K)$ is CA-controllable (with respect to $L(G_i)$, $\Sigma _{uc} \cap \Sigma_i$,
and $\Sigma ^a_c \cap \Sigma_i$) and CA-observable (with respect to $L(G_i)$, $\Sigma _{o} \cap \Sigma_i$, and $\Phi _i^a$), $i=1,2$, then we can design  supervisors $\cS_i$ such that $L_a(\cS_i^a/G_i) = P_i(K)$ by using the following procedure. Such supervisors are resilient to sensor and actuator attacks.

Without loss of generality, we assume that $P_i(K)$ is generated by a sub-automaton $H_i$ of $G_i$. In other words, $P_i(K) = L(H_i)$ for some
\begin{align*}
	H_i = (Q_{H,i}, \Sigma_i, \delta_{H,i}, q_{i,o}),
\end{align*}
where $Q_{H,i} \subseteq Q_i$ and $\delta_{H,i} = \delta|_{Q_{H,i} \times \Sigma_i} \subseteq \delta_i$.

\emph{Step 1.} Construct the extended automaton $G_i^a$ in the same way as the construction of $G^a$ in Equation
(\ref{eq:Ga})
\begin{align*}
	G_i^a & =(Q_i^a, \Sigma_i , \delta _i^a , q_{i,o}, Q_{i,m}^a) .
\end{align*}

\emph{Step 2.}
Extend $H_i$ to $H_i^a$ in the similar way as the extension of $G_i$ to $G_i^a$:
\begin{align*}
	H_i^a & =( Q_{H,i}^a, \Sigma _i, \delta_{H,i} ^a , q_{i,o}, Q^a_{H,i,m}) .
\end{align*}
Note that $H_i^a$ is sub-automaton of $G_i^a$ and $\Theta_i ^a(L(H_i)) =L _m(H_i^a)$.

Replace unobservable transitions in $H_i^a$ by $\varepsilon$-transitions as in Equation (\ref{eq:Gaeps}) and denote the resulting automaton as
\begin{equation}
H_{i,\varepsilon}^a=( Q_{H,i}^a, \Sigma_{i,o} \cup \{ \varepsilon \}, \delta_{H,i,\varepsilon} ^a , q_{i,o}, Q^a_{H,i,m}) ,
\end{equation}
After observing a string $w \in \Phi _i^a (L (H_i))$, the set of all possible states that $H_i$ may be in is called the state estimate which is defined as
\begin{align*}
	E_{H,i}^a(w)= \{& q \in Q_{H,i} : (\exists s \in L(H_i))
	 w \in \Phi _i^a (s) \wedge \delta_{H,i} (q_{i,o}, s)=q \}.
\end{align*}

\emph{Step 3.}  To obtain the state estimates, we convert $H_{i,\varepsilon}^a$ to a deterministic automaton $H_{i,obs}^a$, called CA-observer, as
\begin{align*}
	H_{i,obs}^a & =(X_i, \Sigma_{i,o}, \xi_i, x_{i,o}, X_{i,m})
	= Ac(2^{Q_{H,i}^a}, \Sigma_{i,o}, \xi_i,  UR(\{ q_{i,o} \}), X_{i,m}) ,
\end{align*}
where $Ac(.)$ denotes the accessible part; $UR(.)$ is the
unobservable reach defined, for $x \in X_i$ (that is, $x \subseteq Q_{H,i}^a$), as
$$
UR(x) = \{ q \in Q_{H,i}^a: (\exists q' \in x) q \in \delta _{H,i,\varepsilon}^a (q', \varepsilon) \}.
$$
The transition function $\xi_i$ is defined, for $x \in X_i$ and
$\sigma \in \Sigma _{i,o}$ as
\begin{align*}
	\xi_i (x, \sigma) =UR(\{q \in Q_{H,i}^a: (\exists q' \in x)q \in \delta _{H,i,\varepsilon}^a (q',\sigma) \}).
\end{align*}
The marked states are defined as
$$
X_{i,m} = \{x \in X_i: x \cap Q_{H,i} \not= \emptyset \}.
$$
It is well-known that $L(H_{i,obs}^a)=L(H_{i,\varepsilon}^a)$ and
$L_m(H_{i,obs}^a)=L_m(H_{i,\varepsilon}^a)$.

Using CA-observer $H_{i,obs}^a$, we can calculate the state estimate for any observation $w \in \Phi _i^a (L(H_i))$ as shown in the following theorem.

\begin{theorem} \label{theorem2} 	
After observing $w \in \Phi _i^a (L(H_i))$, the state estimate $E_{H,i}^a(w)$ is given by
$$
E_{H,i}^a(w) = \xi_i (x_{i,o}, w) \cap Q_{H,i}.
$$
\end{theorem}
\noindent {\em Proof:}
The proof is similar to that of Theorem 2 in \cite{zheng2023modeling}.
\hfill \bull

\emph{Step 4.}  To ensure that the supervised system never enters illegal/unsafe states $Q_i \setminus Q_{H,i}$, the set of events that need to be disabled after observing $w \in \Phi _i^a (L(H_i))$ is given by
\begin{align*}
	 \rho (E_{H,i}^a(w))
	=  \{ \sigma \in \Sigma _i \cap \Sigma_c: (\exists q \in E_{H,i}^a(w)) \delta_i (q,\sigma)\in Q_i \setminus Q_{H,i} \}.
\end{align*}
In other words, an event will be disabled if there is  another state in the estimate from which this event takes the system to an illegal/unsafe state.
A state-estimate-based supervisor $\cS_{i,CA}$, called CA-supervisor, can then be designed as
\begin{equation} \label{Equation8}
	\begin{split}
		 \cS_{i,CA} (w)
		= \begin{cases}
			(\Sigma _i \setminus \rho (E_{H,i}^{a}(w)))\cup (\Sigma _{uc} \cap \Sigma _i), &\text{if $ w \in \Phi _i^a (L(H_i)$} \\
			\Sigma _{uc} \cap \Sigma _i, &\text{otherwise}.
		\end{cases}
	\end{split}
\end{equation}
In other words, a local CA-supervisor will enable all local events not leading
 to an illegal/unsafe state, plus all uncontrollable local events. The following theorem can then be proved.

\begin{theorem} \label{theoreme3} 	
	Consider a discrete event system $G_i,i=1,2$ under joint sensor and actuator attacks. For a nonempty closed language $P_i(K) \subseteq L(G_i)$, if $P_i(K)$ is CA-controllable (with respect to $L(G_i)$, $\Sigma _{uc} \cap \Sigma_i$, and $\Sigma ^a_c \cap \Sigma_i$) and CA-observable (with respect to $L(G_i)$, $\Sigma _{o} \cap \Sigma_i$, and $\Phi _i^a$), then the CA-supervisors $\cS_{i,CA}$ defined in Equation (\ref{Equation8}) achieve $P_i(K)$, that is,
	$$
	L _a(\cS^a_{i,CA}/G_i)=P_i(K).
	$$
\end{theorem}

\noindent {\em Proof:} In Theorem 3 of \cite{zheng2023modeling}, a global version of the above result is proved as follows. Consider a discrete event system $G$ under joint sensor and actuator attacks. For a nonempty closed language $K \subseteq L(G)$, if $K$ is CA-controllable and CA-observable, then
$$
L _a(\cS^a_{CA}/G)=K.
$$
The proof for the local version presented here is similar to that of Theorem 3 in \cite{zheng2023modeling}.
\hfill \bull

By Theorem \ref{LGS12}, local supervisors $\cS^a_{i,CA}, i=1,2$ are supervisors for coordination control that achieve $K$, that is,
$$
L_a((\cS_{1,CA}^a \wedge \cS_{2,CA}^a)/G)=K.
$$

\section{Coordination Control under Cyber Attacks without CA-Controllability and CA-Observability}

If $P_i(K)$ is not CA-controllable and/or CA-observable, $i=1,2$,
then there may not exist supervisors $\cS_1$ and $\cS_2$ such that
\begin{align*}
	 L_a((\cS_1^a \wedge \cS_2^a)/G)=K
	 \wedge L_a(\cS_1^a/G_1) \subseteq P_1 (K) \wedge L_a(\cS_2^a/G_2) \subseteq P_2 (K) .
\end{align*}
In this section, we investigate how to design supervisors $\cS^\uparrow_1$ and $\cS^\uparrow_2$ such that
\begin{align*}
	& L_a((\cS_1^{\uparrow, a} \wedge \cS_2^{\uparrow, a})/G) \subseteq K
	 \wedge L_a(\cS_1^{\uparrow, a}/G_1) \subseteq P_1 (K) \wedge L_a(\cS_2^{\uparrow, a}/G_2) \subseteq P_2 (K) .
\end{align*}
The supervisors that satisfy the above condition can ensure safety of the supervised system.

\begin{proposition} \label{Theorem6}
	If $K$ is conditionally decomposable, then for any supervisors $\cS _i$ for $G_i$, $i=1,2$, we have
	\begin{align*}
		& L_a(\cS_1^a/G_1) \subseteq P_1 (K) \wedge L_a(\cS_2^a/G_2) \subseteq P_2 (K) 		\Rightarrow
		 L_a((\cS_1^a \wedge \cS_2^a)/G) \subseteq K .
	\end{align*}
	
\end{proposition}
\noindent {\em Proof:} By Theorem \ref{LGS12},
\begin{align*}
	& L_a((\cS_1^a \wedge \cS_2^a)/G) \\
	= & L_a(\cS_1^a/G_1) \| L_a(\cS_2^a/G_2)\\
	\subseteq & P_1 (K) \| P_2 (K) \\
	= & K
	\mbox{\ \ \ \ } (\mbox{because $K$ is conditionally decomposable})
\end{align*}
\hfill \bull

By Proposition \ref{Theorem6}, for coordination control under cyber attacks, if $K$ is conditionally decomposable but $P_i(K)$ is not CA-controllable and/or CA-observable, $i=1,2$, then we need to  design supervisors $\cS^\uparrow_i, i=1,2$ such that
\begin{align*}
	& L_a(\cS_i^{\uparrow, a}/G_i) \subseteq P_i (K) .
\end{align*}

This can be done using the following procedure.

\emph{Step 1.} Remove states in $H_i$ that can reach illegal states $Q_i \setminus Q_{H,i}$ via a string of uncontrollable and/or attackable events in $G_i$. Denote the resulting automaton by $H_i^\uparrow$. In other words, let
\begin{align*}
	Q_{H,i}^\uparrow = & Q_{H,i} \setminus \{q \in Q_{H,i} : (\exists s \in ( (\Sigma _{uc} \cup \Sigma ^a_c) \cap \Sigma_i )^*)  \delta _i (q, s) \in Q_i \setminus Q_{H,i} \} \\
	H_i^\uparrow = & (Q_{H,i}^\uparrow, \Sigma_i, \delta_{H,i}^\uparrow, q_{i,o}),
\end{align*}
where $\delta_{H,i}^\uparrow = \delta|_{Q_{H,i}^\uparrow \times \Sigma_i} \subseteq \delta_i$. It is well-known \cite{cassandras2021introduction} that $H_i^\uparrow $ generates the supremal controllable sublanguage of $P_i(K)$ with respect to $L(G_i)$ and $(\Sigma _{uc} \cup \Sigma ^a_c) \cap \Sigma_i$, which is equal to the supremal CA-controllable sublanguage of $P_i(K)$ with respect to $L(G_i)$, $\Sigma _{uc} \cap \Sigma_i$, and $\Sigma ^a_c \cap \Sigma_i$, denoted by $P_i(K)^\uparrow$. In other words,
$$
L(H_i^\uparrow) = P_i(K)^\uparrow.
$$
\emph{Step 2.} Extend automaton $G_i$ to
\begin{align*}
	G_i^a & =(Q_i^a, \Sigma_i , \delta _i^a , q_{i,o}, Q_{i,m}^a) .
\end{align*}
\emph{Step 3.} Extend automaton $H_i^\uparrow$ to
\begin{align*}
	H_i^{\uparrow, a} & =( Q_{H,i}^{\uparrow, a}, \Sigma _i, \delta_{H,i} ^{\uparrow, a} , q_{i,o}, Q^{\uparrow, a}_{H,i,m}) .
\end{align*}
Replace unobservable transitions in $H_i^{\uparrow, a}$ by $\varepsilon$-transitions to obtain
$$
H_{i,\varepsilon}^{\uparrow, a}=( Q_{H,i}^{\uparrow, a}, \Sigma_{i,o}, \delta_{H,i,\varepsilon} ^{\uparrow, a} , q_{i,o}, Q^{\uparrow, a}_{H,i,m}) .
$$
After observing a string $w \in \Phi _i^a (L (H_i^\uparrow)) =L_m(H_{i,\varepsilon}^{\uparrow, a})$, the state estimate is defined as
\begin{align*}
	E_{H,i}^{\uparrow, a}(w)= \{& q \in Q_{H,i}^\uparrow : (\exists s \in L(H_i^\uparrow)) w \in \Phi _i^a (s) \wedge \delta_{H,i} ^{\uparrow, a} (q_{i,o}, s)=q \}.
\end{align*}
\emph{Step 4.}  Construct the CA-observer of $H_{i,\varepsilon} ^{\uparrow, a}$ by determinizing $H_{i,\varepsilon}^{\uparrow, a}$ as
\begin{align*}
	H_{i,obs}^{\uparrow, a} & =(X_i^\uparrow, \Sigma_{i,o}, \xi_i^\uparrow, x_{i,o}, X_{i,m}^\uparrow) .
\end{align*}
\emph{Step 5.} Design a CA-supervisor $\cS_{i,CA}^\uparrow$ as

\begin{equation} \label{Equation9}
	\begin{split}
	& \cS_{i,CA}^\uparrow (w)
		= \begin{cases}
			(\Sigma _i \setminus \rho (E_{H,i}^{\uparrow, a}(w)))\cup (\Sigma _{uc} \cap \Sigma _i), &\text{if $ w \in \Phi _i^a (L(H_i^\uparrow)$} \\
			\Sigma _{uc} \cap \Sigma _i, &\text{otherwise }
\end{cases} \\
	\end{split}
\end{equation}
where, $\rho (E_{H,i}^{\uparrow, a}(w)) =  \{ \sigma \in \Sigma _i \cap \Sigma_c: (\exists q \in E_{H,i}^{\uparrow, a}(w)) \delta_i (q,\sigma)\in Q_i \setminus Q_{H,i}^\uparrow$.
The CA-supervisors $\cS_{i,CA}^\uparrow$ designed in the above procedure have the following property.

\begin{theorem} \label{theoreme6}
	Assume that $K$ is conditionally decomposable. Then the  CA-supervisors $\cS_{i,CA}^\uparrow$ ensure safety of the supervised system under joint sensor and actuator attacks, that is,
	\begin{align*}
		L_a((\cS_{1,CA}^{\uparrow,a} \wedge \cS_{2,CA}^{\uparrow,a}/G) \subseteq K .
	\end{align*}
\end{theorem}

\noindent {\em Proof:} Let us first prove, for $i=1,2$,
\begin{align*}
	L_a(\cS_{i,CA}^{\uparrow,a}/G_i) \subseteq P_i (K)^\uparrow = L(H_i^\uparrow)
\end{align*}
by contradiction.

If $L_a(\cS_{i,CA}^{\uparrow,a}/G_i) \subseteq L(H_i^\uparrow)$ is not true, then
\begin{align*}
	& (\exists s \in \Sigma_i^*) (\exists \sigma \in \Sigma_i) s \in L(H_i^\uparrow) \wedge s \sigma \not\in L(H_i^\uparrow)
	\wedge s \sigma \in L_a(\cS_{i,CA}^{\uparrow,a}/G_i)  \\
	\Leftrightarrow
	& (\exists s \in \Sigma_i^*) (\exists \sigma \in \Sigma_i) s \in L(H_i^\uparrow) \wedge s \sigma \not\in L(H_i^\uparrow)  \wedge s \in L_a(\cS_{i,CA}^{\uparrow,a}/G_i) \wedge s \sigma \in L (G_i) \\
	& \wedge (\sigma \in \Sigma _{uc} \vee (\exists w \in \Phi_i^a (s))(\exists \gamma_i^a \in \cS_{i,CA}^{\uparrow,a}(w)) \sigma \in \gamma_i^a)  \\
	& (\mbox{by the definition of } L_a(\cS_{i,CA}^{\uparrow,a}/G_i) ) \\
	\Leftrightarrow
	& (\exists s \in \Sigma_i^*) (\exists \sigma \in \Sigma_i) s \in L(H_i^\uparrow) \wedge s \sigma \not\in L(H_i^\uparrow)
\wedge s \in L_a(\cS_{i,CA}^{\uparrow,a}/G_i) \wedge s \sigma \in L (G_i) \\
	& \wedge (\sigma \in \Sigma _{uc} \cup \Sigma _c^a \vee (\exists w \in \Phi_i^a (s)) \sigma \in \cS_{i,CA}^{\uparrow} (w)) \\
	& (\mbox{by the definition of } \cS_{i,CA}^{\uparrow,a}(w) ) \\
	\Leftrightarrow
	& (\exists s \in \Sigma_i^*) (\exists \sigma \in \Sigma_i) s \in L(H_i^\uparrow) \wedge s \sigma \not\in L(H_i^\uparrow)
 \wedge s \in L_a(\cS_{i,CA}^{\uparrow,a}/G_i) \wedge s \sigma \in L (G_i) \\
	& \wedge (\sigma \in \Sigma _{uc} \cup \Sigma _c^a \vee (\exists w \in \Phi_i^a (s)) \sigma \not\in \rho (E_{H,i}^{\uparrow, a}(w))) \\
	& (\mbox{by the definition of } \cS_{i,CA}^{\uparrow}(w) ) \\
	\Leftrightarrow
	& (\exists s \in \Sigma_i^*) (\exists \sigma \in \Sigma_i) s \in L(H_i^\uparrow) \wedge s \sigma \not\in L(H_i^\uparrow) \\
	& \wedge s \in L_a(\cS_{i,CA}^{\uparrow,a}/G_i) \wedge s \sigma \in L (G_i)  \wedge (\exists w \in \Phi_i^a (s)) \sigma \not\in \rho (E_{H,i}^{\uparrow, a}(w)) \\
	& (\mbox{because }  s \in L(H_i^\uparrow) \wedge s \sigma \not\in L(H_i^\uparrow) \Rightarrow \sigma \not\in \Sigma _{uc} \cup \Sigma _c^a ) \\
	\Leftrightarrow
	& (\exists s \in \Sigma_i^*) (\exists \sigma \in \Sigma_i) s \in L(H_i^\uparrow) \wedge s \sigma \not\in L(H_i^\uparrow)
	 \wedge s \in L_a(\cS_{i,CA}^{\uparrow,a}/G_i) \wedge s \sigma \in L (G_i) \\
	& \wedge (\exists w \in \Phi_i^a (s)) \neg (\exists q \in E_{H,i}^{\uparrow, a}(w)) \delta_i (q,\sigma)\in Q_i \setminus Q_{H,i}^\uparrow  \\
	& (\mbox{by the definition of } \rho (E_{H,i}^{\uparrow, a}(w)) ) \\
	\Leftrightarrow
	& (\exists s \in \Sigma_i^*) (\exists \sigma \in \Sigma_i) s \in L(H_i^\uparrow) \wedge s \sigma \not\in L(H_i^\uparrow) \\
	& \wedge s \in L_a(\cS_{i,CA}^{\uparrow,a}/G_i) \wedge s \sigma \in L (G_i) \wedge (\exists w \in \Phi_i^a (s))
	 (\forall q \in E_{H,i}^{\uparrow, a}(w)) \delta_i (q,\sigma) \not\in Q_i \setminus Q_{H,i}^\uparrow  \\
	\Leftrightarrow
	& (\exists s \in \Sigma_i^*) (\exists \sigma \in \Sigma_i) s \in L(H_i^\uparrow) \wedge s \sigma \not\in L(H_i^\uparrow)  \wedge s \in L_a(\cS_{i,CA}^{\uparrow,a}/G_i) \\
    & \wedge s \sigma \in L (G_i) \wedge (\exists w \in \Phi_i^a (s)) (\forall s' \in L(H_i^\uparrow)) w \in \Phi _i^a (s') \Rightarrow s' \sigma \not\in L(G_i) \setminus  L(H_i^\uparrow) \\
	\Rightarrow
	& (\exists s \in \Sigma_i^*) (\exists \sigma \in \Sigma_i) s \in L(H_i^\uparrow) \wedge s \sigma \not\in L(H_i^\uparrow) \\
	& \wedge s \in L_a(\cS_{i,CA}^{\uparrow,a}/G_i) \wedge s \sigma \in L (G_i) \wedge (\exists w \in \Phi_i^a (s))  s \sigma \not\in L(G_i) \setminus  L(H_i^\uparrow) \\
	& (\mbox{by letting } s'=s ) \\
	\Leftrightarrow
	& (\exists s \in \Sigma_i^*) (\exists \sigma \in \Sigma_i) s \in L(H_i^\uparrow) \wedge s \sigma \not\in L(H_i^\uparrow) \\
	& \wedge s \in L_a(\cS_{i,CA}^{\uparrow,a}/G_i) \wedge s \sigma \in L (G_i)  \wedge (s \sigma \not\in L(G_i) \vee s \sigma \in L(H_i^\uparrow) ) .
\end{align*}
Clearly, this is a contradiction.

Now, let us prove the theorem as follows.
\begin{align*}
	& L_a(\cS_{1,CA}^{\uparrow,a}/G_1) \subseteq P_1 (K)^\uparrow \wedge L_a(\cS_{2,CA}^{\uparrow,a}/G_2) \subseteq P_2 (K)^\uparrow \\
	\Rightarrow
	& L_a(\cS_{1,CA}^{\uparrow,a}/G_1) \subseteq P_1 (K) \wedge L_a(\cS_{2,CA}^{\uparrow,a}/G_2) \subseteq P_2 (K) \\
	\Rightarrow
	& L_a((\cS_{1,CA}^{\uparrow,a} \wedge \cS_{2,CA}^{\uparrow,a}/G) \subseteq K  \mbox{\ \ \ \ }(\mbox{by Proposition \ref{Theorem6}}) .
\end{align*}
\hfill \bull

\section{Illustrative Examples}

In this section, we validate our approach against both stealthy and non-stealthy attacks. Furthermore, we demonstrate the empirical results of this work by a case of autonomous vehicles.

\subsection{A comparative study on stealthy and non-stealthy attacks}

	Our approach does not depend on detecting attacks and hence works for both stealthy and non-stealthy attacks. In fact, if the supervisor knows that the attacker is stealthy, then it gives the supervisor more information about the attacker, which lead to a more permissive closed-loop. We demonstrate this via the subsequent example.
	
\begin{figure}[h!]
    \centering
    \begin{subfigure}[b]{0.45\textwidth}
        \centering
        \includegraphics[scale=0.5]{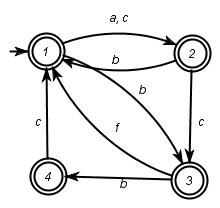}
        \caption{The automaton $G_1$.}
        \label{fig:AG1}
    \end{subfigure}
    \begin{subfigure}[b]{0.45\textwidth}
        \centering
        \includegraphics[scale=0.5]{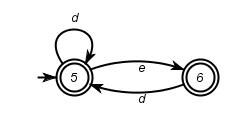}
        \caption{The automaton $G_2$.}
        \label{fig:AG2}
    \end{subfigure}
    \caption{Two automata $G_1$ and $G_2$.}
    \label{fig:AG12}
    \end{figure}

	Let $n=2$. Consider $G_1$ shown in Fig. \ref{fig:AG1} and $G_2$ shown in Fig. \ref{fig:AG2} with $\Sigma_1 = \{a, b, c, f\}$ and $\Sigma_2 = \{d, e\}$.

%
	
	The overall system $G$ is the synchronous product of $G_1$ and $G_2$:
	$$
	G = G_1 \| G_2.
	$$
	
	\begin{figure}[h!]
		\centering
		\includegraphics[scale=0.5]{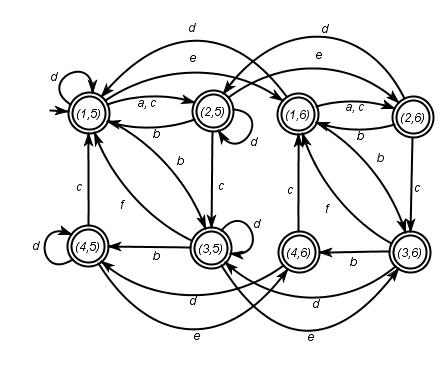}
		\caption{The automaton $G = G_1 \| G_2$. }
		\label{fig:FB}
	\end{figure}	
	
\begin{figure}[h!]
    \centering
    \begin{subfigure}[b]{0.45\textwidth}
        \centering
        \includegraphics[scale=0.5]{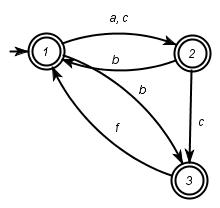}
        \caption{The automaton $H_1$ for $K_1$.}
        \label{fig:AH1}
    \end{subfigure}
    \begin{subfigure}[b]{0.45\textwidth}
        \centering
        \includegraphics[scale=0.5]{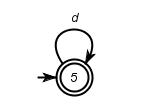}
        \caption{The automaton $H_2$ for $K_2$.}
        \label{fig:AH2}
    \end{subfigure}
    \caption{Two automata $H_1$ and $H_2$.}
    \label{fig:AH12}
    \end{figure}

\begin{figure}[h!]
		\centering
		\includegraphics[scale=0.5]{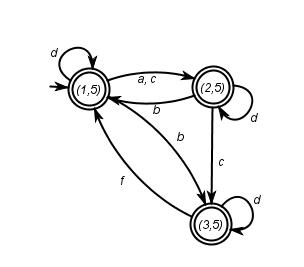}
		\caption{The automaton $H$ for $K = K_1 \| K_2$. }
		\label{fig:AK}
	\end{figure}

	Let $K_1 = L(H_1)$, where $H_1$ is shown in Fig. \ref{fig:AH1}. Let $K_2 = L(H_2)$, where $H_2$ is shown in Fig. \ref{fig:AH2}.  Let the specification language $K$ be the synchronous product of $K_1$ and $K_2$:
	$$
	K = K_1 \| K_2.
	$$
	The automaton $H$ for $K$ is shown in Fig. \ref{fig:AK}. Clearly, the automaton $H_i$ is a sub-automaton of $G_i$, $i=1,2$.

%
	
	Since $P_1 (K) = K_1$ and $P_2 (K) = K_2$, $K = P_1(K) \|  P_2 (K)$, that is, $K$ is conditionally decomposable with respect $\Sigma _1$ and $\Sigma _2$.
	
	To focus on attacks, assume that all events are controllable and observable, that is, $\Sigma = \Sigma_{c} = \Sigma_{o}$. Then all languages are controllable and observable. Without attacks, local supervisors $\cS_i$ can be easily obtained based on $G_i$: $\cS_1$ disables $b$ at state 3 of $G_1$ and $\cS_2$ disables $e$ at state 5 of $G_2$.
	
	Let us consider two cases of attacks: One for non-stealthy attacks and the other for stealthy attacks.
	
	\vspace{0.3cm}
	
	\emph{Case 1. Non-stealthy attacks:} Let us first consider the following non-stealthy sensor attack:
	\begin{align*}
		& \Sigma _o ^a = \{a, b, f\}  \\
		& A_{a} = \{c\}, \ \ \ \ \ A_{b} = \{b, c\}, \ \ \ \ \ A_{f} = \{f, c\} .
	\end{align*}

\begin{figure}[h!]
		\centering
		\includegraphics[scale=0.5]{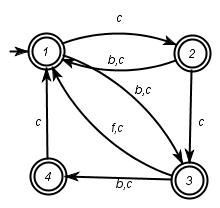}
		\caption{The automaton $G_1^a$ for Case 1. }
		\label{fig:AG1a}
	\end{figure}

\begin{figure}[h!]
		\centering
		\includegraphics[scale=0.5]{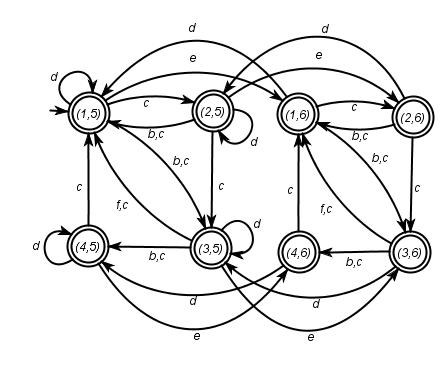}
		\caption{The automaton $G^a = G_1^a \| G_2$ for Case 1. }
		\label{fig:AGa}
	\end{figure}

	The corresponding $G_1^a$ and $G^a = G_1^a \| G_2$ are shown in Fig. \ref{fig:AG1a} and Fig. \ref{fig:AGa}, respectively.
	Note that, since there are no sensor attacks on $G_2$, $G_2^a = G_2$.
	
	Since $L(G_1^a) \not\subseteq L(G_1)$ and $L(G^a) \not\subseteq L(G)$, the attack is not stealthy. For example, string $ccc \in L(G_1^a)$ but $ccc \not\in L(G_1)$. Furthermore, $P_1(K) = K_1$ is not CA-observable and no supervisor $\tcS_1$ exists to achieve $K_1$. For example, after observing $cc$, $G_1$ can be in either state 1 or state 3, while in state 1, $b$ needs to be enabled, in state 3, $b$ needs to be disabled. Thus the closed-loop under attack will be restrictive.

	
	\vspace{0.3cm}
	
\begin{figure}[h!]
		\centering
		\includegraphics[scale=0.5]{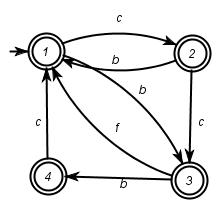}
		\caption{The automaton $G_1^a$ for Case 2. }
		\label{fig:AG1a2}
	\end{figure}

\begin{figure}[h!]
		\centering
		\includegraphics[scale=0.5]{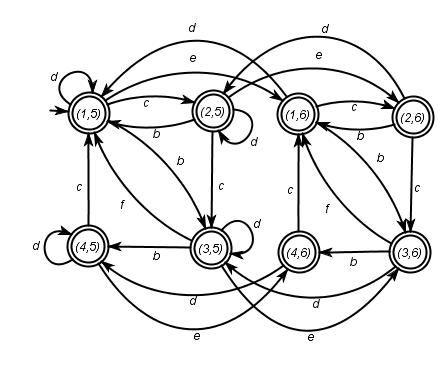}
		\caption{The automaton $G^a = G_1^a \| G_2$ for Case 2. }
		\label{fig:AGa2}
	\end{figure}

	\emph{Case 2. Stealthy attacks:} If the supervisor knows that the attacks are stealthy, then after observing $b$ or $f$, the attacker cannot change $b$ or $f$ to $c$. In other words, we have,
	\begin{align*}
		& \Sigma _o ^a = \{a, b, f\}  \\
		& A_{a} = \{c\}, \ \ \ \ \ A_{b} = \{b\}, \ \ \ \ \ A_{f} = \{f\} .
	\end{align*}
	The corresponding $G_1^a$ and $G^a = G_1^a \| G_2$ are shown in Fig. \ref{fig:AG1a2} and Fig. \ref{fig:AGa2}, respectively.
	
	Since $L(G_1^a) \subseteq L(G_1)$ and $L(G^a) \subseteq L(G)$, the attack is stealthy.	
	It can be checked that $P_i(K) = K_i$ is CA-controllable and  CA-observable. Therefore, there exist local supervisors $\tcS_1$ and $\tcS_2$ such that $L_a((\tcS_1^a \wedge \tcS_2^a)/G)=K \wedge L_a(\tcS_i^a/G_i) \subseteq P_i (K) , i=1,2$.
	
	The local supervisor $\tcS_1$ can be obtained based on $G_1^a$: $\tcS_1$ disables $b$ at state 3 of $G_1^a$. This supervisor is robust under sensor attacks. On the other hand,  since there are no sensor attacks on $G_2$, $\tcS_2 = \cS_2$.
	
	Note that event $a$ does not appear in $G_1^a$. This does not change the fact that $L_a(\tcS_1^a/G_1) = P_1 (K) = K_1$ and $L_a((\tcS_1^a \wedge \tcS_2^a)/G) = K$.
	
	\vspace{0.3cm}	
	
	This example shows that our approach still works even if the attacks are stealthy. Furthermore, if the attacks are stealthy, then the attacked model can be refined, which will make the resilient supervisor easier to obtain, resulting in a more permissive closed-loop of the attacked systems, while still keeping the safety requirement.
	
	Note that our approach does not require the attacks be stealthy. The design procedure is the same for both stealthy and non-stealthy attacks. This is in contrast to other approaches proposed in the literature and makes our approach different and innovative.

\subsection{A comprehensive example on autonomous vehicles}

\begin{figure}[htb]
	\centering
	\includegraphics[scale=0.4]{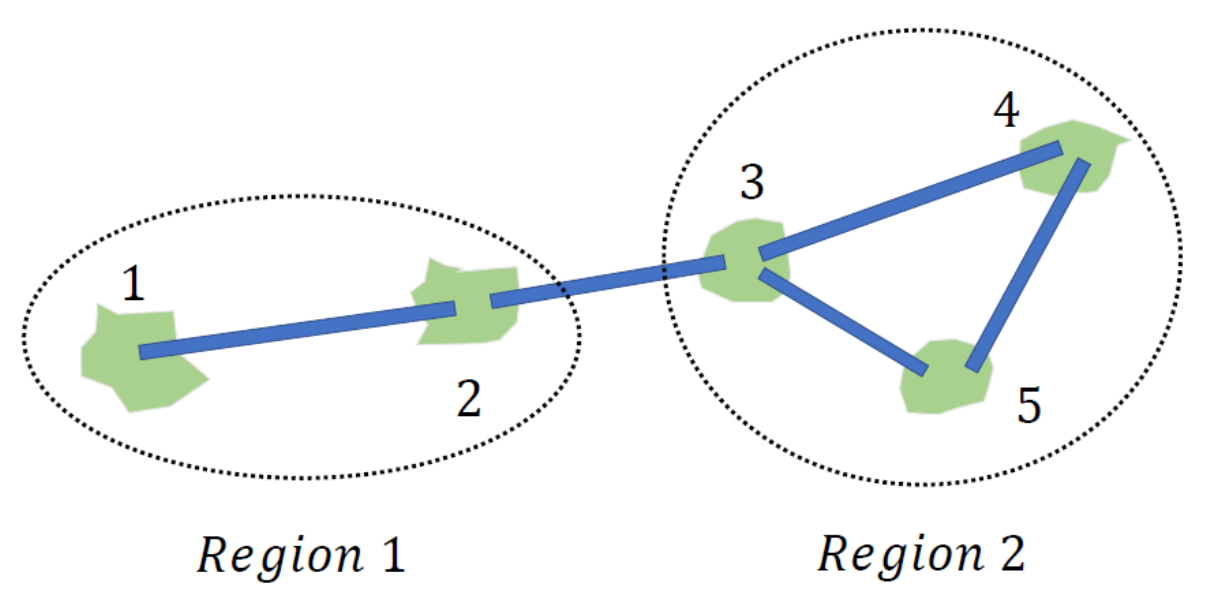}
	\caption{Five islands connected by five bridges and divided into two regions. }
	\label{fig:Island}
\end{figure}

Consider five islands connected by five bridges as shown in Fig. \ref{fig:Island}. Five islands are divided into two regions. Region 1 consists of Islands 1 and 2. Region 2 consists of Islands 3, 4, and 5.

\begin{figure}[htb]
	\centering
	\includegraphics[scale=0.45]{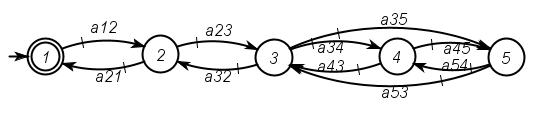}
	\includegraphics[scale=0.45]{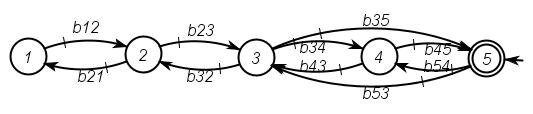}
	\caption{Automata $G_a$ (top) and $G_b$ (bottom). }
	\label{fig:G1G2}
\end{figure}

Two autonomous vehicles $V_a$ and $V_b$ are moving among islands to perform some tasks (not specified). Initially, $V_a$ is on Island 1 and $V_b$ is on Island 5.
The automata modeling the movement of the vehicles are shown in Fig. \ref{fig:G1G2}.

In the figure, states of $G_a$ are denoted by $i$, where $i=1,2,3,4,5$, corresponding to the vehicle $V_a$ is on Island $i$.
Events are $aij$ - vehicle $V_a$ moves from Island $i$ to Island $j$. Event $aij$ is defined if there is a bridge from Island $i$ to Island $j$. Controllable events are denoted in the figures by small marks on the edges/arrows (that is, $\not\longrightarrow$).

Automaton $G_b$ is similarly defined.

\begin{figure}[htb]
	\centering
	\includegraphics[scale=0.4]{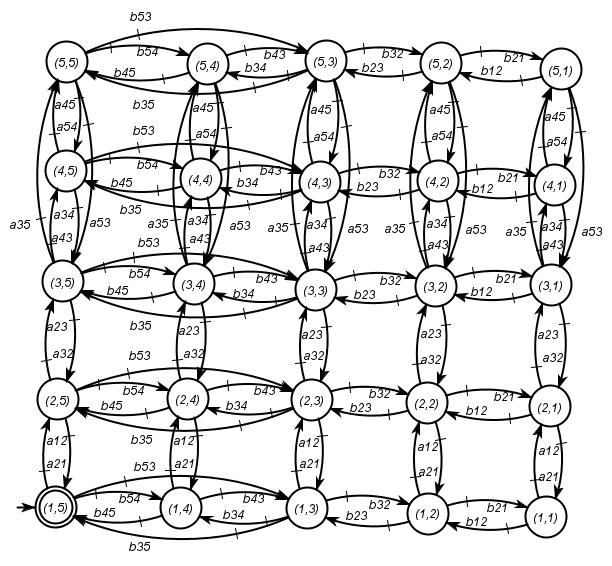}
	\caption{Automaton $G$ of the system. }
	\label{fig:G}
\end{figure}

The overall system is obtained by taking the synchronous product of $G_a$ and $G_b$:
$$
G = G_a \| G_b.
$$
$G$ is shown in Fig. \ref{fig:G}. The event set of $G$ is
$$
\Sigma = \{ aij, bij : \mbox{there is a bridge from Island $i$ to Island $j$}\}.
$$
We assume that all events are observable, that is, $\Sigma _o = \Sigma$.

\begin{figure}[htb]
	\centering
	\includegraphics[scale=0.4]{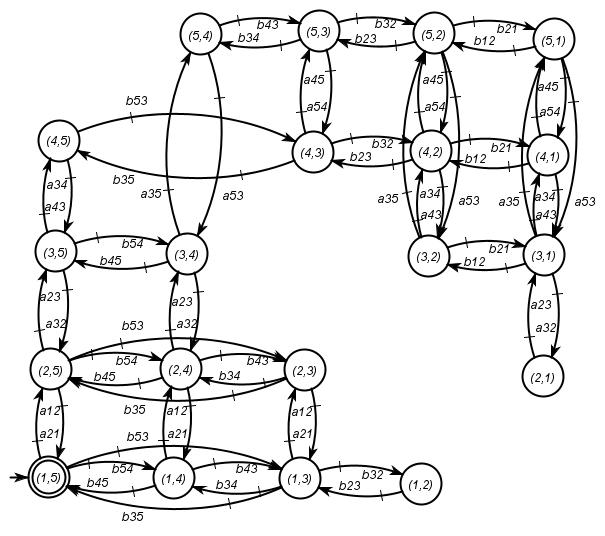}
	\caption{The specification automaton $H$. }
	\label{fig:H}
\end{figure}

To avoid conflicts between two vehicles, we would like to control the bridges so that two vehicles will never be on the same island at the same time. Therefore, the safety specification described by automaton $H$ shown in Fig. \ref{fig:H} is obtained by removing states labeled (1,1), (2,2), (3,3), (4,4), and (5,5) from automaton $G$. The corresponding language is $K=L(H)$.

We assume that an attacker can attack the sensors on the bridge between Island 2 and Island 3 as well as the actuators on the bridge between Island 1 and Island 2. In other words,
\begin{align*}
	\Sigma _o ^a & = \{a23, a32, b23, b32 \}, \  \Sigma _c ^a = \{a12, a21, b12, b21 \} .
\end{align*}
Let the sensor attacks be static and specified by
\begin{equation} \label{Equation10}
	\begin{split}
		A_{a23} = \{a23, b23 \}, \  A_{a32} = \{ a12 \},
	\ A_{b23} = \{b23, b12 \}, \  A_{b32} = \{ a32 \} .
	\end{split}
\end{equation}
We will use coordination control to design two local supervisors $\cS_1$ and $\cS_2$ to control the plant $G$. Supervisor $\cS_1$ controls Region 1 consisting of Islands 1 and 2. Supervisor $\cS_2$ controls Region 2 consisting of Islands 3, 4, and 5.
We choose the corresponding event sets such that the region border crossing events, namely $a23, a32, b23$, and $b32$ are added to events in both regions 1 and 2. Therefore,
\begin{align*}
	\Sigma _1 = & \{ a12, a21, a23, a32, b12, b21, b23, b32 \}, \\
	\Sigma _2 = & \{ a23, a32, a34, a43, a45, a54, a35, a53,
	 b23, b32, b34, b43, b45, b54, b35, b53 \} .
\end{align*}

\begin{figure}[htb]
	\centering
	\includegraphics[scale=0.99]{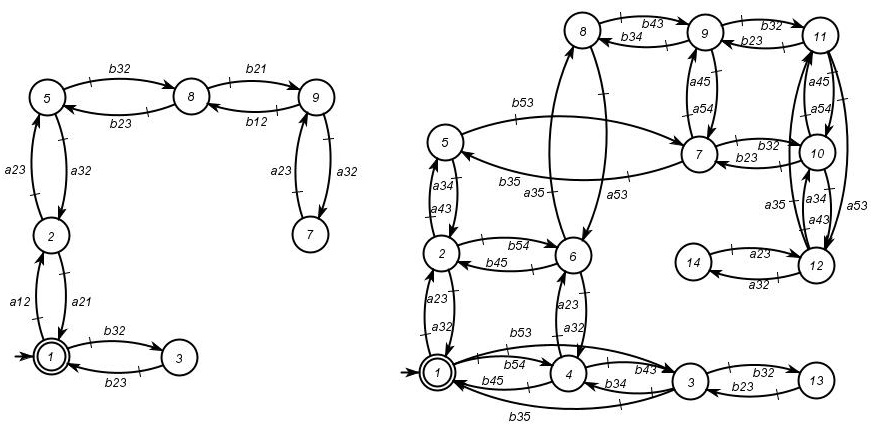}
	\caption{Automaton $H_1$ for $P_1(K)$ (top) and automaton $H_2$ for $P_2 (K)$ (bottom). }
	\label{fig:K1K2}
\end{figure}

Let us check if $K$ is conditionally decomposable with respect to $\Sigma _1$ and $\Sigma _2$. Automaton $H_1$ for $P_1(K)$ and automaton $H_2$ for $P_2 (K)$ are calculated as shown in Fig. \ref{fig:K1K2}. It can be checked that $K = P_1(K) \| P_2 (K)$. Hence, $K$ is conditionally decomposable.

\begin{figure}[htb]
	\centering
	\includegraphics[scale=0.45]{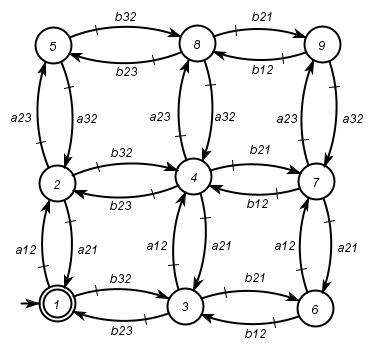}
	\caption{Automaton $G_1$. }
	\label{fig:G1}
\end{figure}

To design local supervisor $\cS_1$, we obtain automaton $G_1$ for $P_1(L(G))$ as shown in Fig. \ref{fig:G1}. Note that automaton $H_1$ in Fig. \ref{fig:K1K2} is a sub-automaton of $G_1$.

It can be checked that $P_1(K)$ is not CA-controllable with respect to $L(G_1)$, $\Sigma _{uc} \cap \Sigma_1$, and $\Sigma ^a_c \cap \Sigma_1$.
Therefore, let us use method in Section 5 to design supervisors $\cS^\uparrow_1$ and $\cS^\uparrow_2$ such that
\begin{align*}
	L_a(\cS_1^{\uparrow, a}/G_1) \subseteq P_1 (K) \wedge L_a(\cS_2^{\uparrow, a}/G_2) \subseteq P_2 (K) .
\end{align*}
Since $K$ is conditionally decomposable, by Proposition \ref{Theorem6}, we have
\begin{align*}
	& L_a((\cS_1^{\uparrow, a} \wedge \cS_2^{\uparrow, a})/G) \subseteq K .
\end{align*}

Such supervisors can be designed using Theorem \ref{theoreme6} as follows.

\begin{figure}[htb]
	\centering
	\includegraphics[scale=0.45]{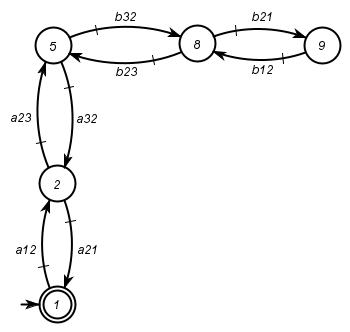}
	\caption{Autumaton $H_1^\uparrow = (Q_{H,1}^\uparrow, \Sigma_1, \delta_{H,1}^\uparrow, q_{1,o})$.}
	\label{fig:H1up}
\end{figure}

We first obtain $H_1^\uparrow = (Q_{H,1}^\uparrow, \Sigma_1, \delta_{H,1}^\uparrow, q_{1,o})$ by removing states 3 and 7 in $H_1$, because $b21 \in \Sigma _c ^a$ can take $G$ from state 3 to illegal state 6 and $a21 \in \Sigma _c ^a$ can take $G$ from state 7 to illegal state 6. The resulting automaton is shown in Fig. \ref{fig:H1up}.

\begin{figure}[htb]
	\centering
	\includegraphics[scale=0.45]{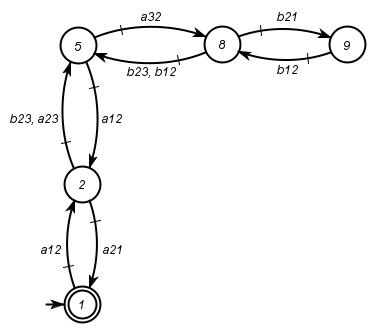}
	\caption{CA-observer $H_{1,obs}^{\uparrow, a} =(X_1^\uparrow, \Sigma_{1,o}, \xi_1^\uparrow, x_{1,o}, X_{1,m}^\uparrow)$.}
	\label{fig:H1obs}
\end{figure}

Since all events are observable and the attacker can attack the sensors on the bridge between Island 2 and Island 3, we can obtain CA-observer $H_{1,obs}^{\uparrow, a}$ as shown in Fig. \ref{fig:H1obs} based on attack languages in Equation (\ref{Equation10}).

Based on $H_{1,obs}^a$, the CA-supervisor $\cS_{1,CA}^\uparrow$ can be designed using Equation (\ref{Equation9}). After observing $w \in \Phi _1^a (L(H_1)$, if the corresponding state $\xi_1 (x_{1,o}, w)$ is $1$ or $2$, then $b32$ is disabled; if the corresponding state $\xi_1 (x_{1,o}, w)$ is $8$ or $9$, then $a32$ is disabled.

The design of supervisor $\cS_{2,CA}^\uparrow$ is similar to that of supervisor $\cS_{1,CA}^\uparrow$ and is hence omitted.


\section{Conclusion}

In this paper, we investigate supervisory control of discrete event systems in the presence of cyber attacks targeting both sensors and actuators. Our focus lies within the framework of coordination control, where we employ local supervisors to address this challenge.
In a future work we plan to develop hierarchical supervisory control under cyber attacks and to study equality between closed-loop achieved by the monolithic supervisor and the closed-loop achieved by the joint action of  modular/coordination supervisors in the case, where CA-controllability and CA-observability are not satisfied based on the results from hierarchical control.

\bibliographystyle{plain}
\bibliography{CoordC_R4arXiv}             

\end{document}